\documentclass[letterpaper]{article} %
\usepackage[preprint]{aaai2027}  %
\usepackage[hyphens]{url}  %
\usepackage{graphicx} %
\usepackage{natbib}  %
\usepackage{caption} %
\usepackage{algorithm}
\usepackage{algorithmic}

\usepackage{newfloat}
\usepackage{listings}
\DeclareCaptionStyle{ruled}{labelfont=normalfont,labelsep=colon,strut=off} %
\floatstyle{ruled}
\newfloat{listing}{tb}{lst}{}
\floatname{listing}{Listing}

\usepackage{times}
\usepackage{soul}
\usepackage{url}
\usepackage{subcaption}
\usepackage{graphicx}
\usepackage{booktabs}
\usepackage{algorithm}
\usepackage{algorithmic}
\usepackage[switch]{lineno}

\usepackage{amssymb}
\usepackage{mathtools}
\usepackage{nicefrac}
\usepackage{paralist}

\usepackage[shortlabels,inline]{enumitem}

\usepackage[nameinlink]{cleveref}
\crefname{problem}{problem}{problems} 
\Crefname{problem}{Problem}{Problems} 
\Crefname{assumption}{Assumption}{Assumptions} 

\usepackage[appendix=append, forwardlinking=yes, bibliography=common]{apxproof} 

\newtheoremrep{theorem}{Theorem}
\newtheoremrep{lemma}{Lemma}
\newtheoremrep{proposition}{Proposition}
\newtheoremrep{corollary}{Corollary}
\newtheoremrep{assumption}{Assumption}

\theoremstyle{definition}
\newtheorem{definition}{Definition}[section]
\newtheorem*{definition*}{Definition}

\newtheorem*{theorem*}{Theorem}

\newtheorem*{problem*}{Problem}

\newcommand{\eps}{\varepsilon}

\newcommand{\R}[0]{\mathbb{R}}

\newcommand{\SW}[0]{\mathrm{SW}}

\newcommand{\game}[0]{G}
\renewcommand{\u}[0]{\textbf{u}}
\renewcommand{\a}[0]{a}

\newcommand{\C}[0]{\Delta}

\newcommand{\F}[0]{\mathcal{F}}

\newcommand{\red}[1]{\bar{#1}}

\newcommand{\remap}[0]{\Psi}
\newcommand{\remapu}[0]{\hat \Psi}

\newcommand{\iso}[0]{\phi}

\newcommand{\pgeq}[0]{\succeq}
\newcommand{\pge}[0]{\succ}

\newcommand{\tgame}[0]{\mathcal{T}}

\newcommand{\rf}[0]{\remap} %
\newcommand{\redf}[0]{\mathtt{Red}} %

\newcommand{\fmb}[0]{\mathcal{F}^{\mathrm{MB}}} %

\makeatletter
\let\oldabs\abs
\def\abs{\@ifstar{\oldabs}{\oldabs*}}
\let\oldnorm\norm
\def\norm{\@ifstar{\oldnorm}{\oldnorm*}}
\makeatother

\usepackage{ifthen}
\usepackage{xcolor}

\usepackage{verbatim}
\usepackage{csquotes}

\usepackage{array}
\usepackage{booktabs}

\title{Expanding and Evaluating the Applicability of Safe Pareto Improvements}
\author{
    Nathaniel Sauerberg\textsuperscript{\rm 1},
    Caspar Oesterheld\textsuperscript{\rm 2}
}
\affiliations{
    \textsuperscript{\rm 1}University of Texas at Austin\\
    \textsuperscript{\rm 2}Carnegie Mellon University\\
    nsauerberg@utexas.edu,
    oesterheld@cmu.edu
}

\begin{document}

\maketitle

\begin{abstract}
Even when players can make credible joint commitments in games, disagreement about the game's default outcome (e.g., about equilibrium selection) can prevent them from agreeing on any particular payoff profile. We study safe Pareto improvements (SPIs): commitments that sidestep this obstacle by inducing a new game that is isomorphic to the original (after iteratively removing strictly dominated actions) but entrywise Pareto better. We do so by constructing token games: cheap-talk games rendered meaningful by commitments to take payoff-relevant actions (in prior work, strategies in the original game) as a function of the token outcome. Since these token games preserve the original game's strategic structure, players should expect them to be played isomorphically and hence agree that they Pareto improve on the original.

In this work, we study token game SPIs under three additional commitment abilities: (1) burning utility, (2) transferring utility between players, and (3) transferring utility with an outside party conditional on the token outcome (e.g., via insurance-like contracts). We characterize when SPIs exist in each case and give efficient algorithms for optimizing over them. Finally, we propose evaluating SPI frameworks by how often they enable SPIs in randomly generated games. In such experiments, our methods substantially outperform prior work.
\end{abstract}

\section{Introduction}

It's well-known that joint commitment in games can enable mutually beneficial agreements. 
For instance, players can commit to play (Cooperate, Cooperate) in a prisoners' dilemma. 
This is considered mutually beneficial because it's generally accepted that the default outcome would be (Defect, Defect). 
However, in other games, it's much less clear how to evaluate how the game would be played by default, and therefore whether a particular agreement is mutually beneficial. 
For instance, consider the game of Chicken. 
Is an agreement to randomize uniformly over (Dare, Swerve) and (Swerve, Dare) mutually beneficial? 
Answering this requires evaluating what equilibrium, if any, would be played by default. 
For example, if both players subjectively expect to Dare while the other Swerves, neither will consider such an agreement beneficial.

Much prior work has characterized the impact of various commitment frameworks with so-called \textit{folk theorems} 
\cite[e.g.][]{Tennenholtz04:Program, kalai2010commitment, deng2017disarmament, oesterheld2023similarity}: any feasible payoff profile can be achieved in equilibrium so long as each player achieves at least the minimum utility they can unilaterally guarantee for themselves (\enquote{individual rationality}).  
The upside of such results is limited by a general version of the problem discussed above: any such Pareto-optimal payoff profile can be reached, but agreeing on a \textit{particular} such profile might require agreeing how the game would be played by default. 

In this paper, we allow players to make joint, binding commitments before playing a game,
and seek commitments which guarantee a Pareto improvement regardless of how the game is played. 
We call such agreements safe Pareto improvements (SPIs) \cite{SPI-original}. 
In particular, we seek commitments which induce a game which is isomorphic to the original, but Pareto better. 
This approach overcomes several difficulties. %
First, it may just be very unclear how the players would play the game by default (e.g. if and how they would have resolved equilibrium selection), and therefore whether agreeing to play any particular outcome is an improvement on the default.
Second, even if the players do have beliefs about how the game would be played by default, they might disagree about what will happen. %
For instance, each player might expect to achieve their preferred Nash equilibrium in Chicken, in which case no mutually beneficial agreement exists. %
Finally, the incentives around bargaining and strategic posturing might prevent the elicitation of the players' true beliefs. 
In Chicken, for instance, the players might claim to expect their preferred equilibrium regardless of their true beliefs, so as to claim that they intend to Dare and the other player should Swerve. 

Our approach circumvents all three of these potential obstacles for achieving Pareto improvements. 
By inducing a game which is isomorphic to the original, we preserve the strategic structure of the game: the equilibria, the dominance relationships, etc. %
Therefore, each player should plausibly play the new game the same as the original, and have the same beliefs about how the other players will play. 
But because every possible outcome in the new game Pareto-dominates the corresponding outcome in the original game, such commitment leaves all players better off, regardless of their beliefs.

Our approach to achieving these goals leverages token games \cite{Sauerberg2025_promises_EPV-SPIs}.
A token game is a symbolic normal-form game played over cheap talk.
For instance, a token game with action spaces $T_i$ could be played by having each Player $i$ privately write down their action $t_i$ on a piece of paper, and then simultaneously revealing the $t_i$ to determine the token outcome $t$.  
This is not inherently meaningful, but can be rendered meaningful by commitments to act in the original game as a function of the token outcome. 
The original game payoffs of these actions then determine the payoffs of the token game.

As an example, consider \Cref{game:chicken_w_concede_and_payments_token_spi}. 
The action labels stand for Dare, Swerve, and Concede, and T stands for token.
Chicken with Concede is a version of Chicken in which Player $2$ has an additional action, Concede, which is strictly dominated by Swerve. 
Therefore, we assume Concede is never played and can be removed without changing how the game is played. 
We construct a token game (\Cref{game:chicken_w_concede_and_payments_token_spi}, Right) which is isomorphic to Chicken with Concede (excluding Concede), but Pareto better. 

First, observe that the payoffs of the token game can be achieved by committing to play strategies and transfer utility in the original game. 
For example, if the outcome is (TD, TS), the players commit to play (Dare, Concede) in the original game, and Player $1$ commits to transfer $3$ utility to Player $2$, resulting in payoffs of $(8-3, 0+3) = (5, 3)$.
The payoffs of (TS, TD) and (TS, TS) can be implemented by analogous commitments, while the payoffs of (TD, TD) can be implemented by committing to play a correlated strategy profile that randomizes between (D, D) and (S, S). 

Second, observe that the token game is isomorphic to the original game (with Concede removed).
For an outcome in the original game with payoff profile $(v_1, v_2)$, the corresponding outcome in the token game has payoff profile $(0.5 v_1 + 2.5, 0.5 v_2 + 2.5)$.
We'll therefore assume that the token game would be played isomorphically to the original game, and that the players' beliefs about how it would be played are similarly isomorphic. %
But each outcome in the token game Pareto-dominates the corresponding outcome in the original game. 
Therefore, regardless of their beliefs about how the original game would be played, both players should prefer to play the token game, and regardless of how the original game is played, the token game guarantees a Pareto improvement on the original. 
We've found our desired {SPI}.

\begin{table}[t]
\centering
\begin{minipage}[t]{0.48\columnwidth}
\centering
\begin{tabular}{cccc}
& D & S & C \\ \cline{2-4} 
\multicolumn{1}{c|}{D} & \multicolumn{1}{c|}{-9, -9} & \multicolumn{1}{c|}{5, 1} & \multicolumn{1}{c|}{8, 0} \\ \cline{2-4} 
\multicolumn{1}{c|}{S} & \multicolumn{1}{c|}{1, 5} & \multicolumn{1}{c|}{3, 3} & \multicolumn{1}{c|}{8, 0} \\ \cline{2-4} 
\end{tabular}
\end{minipage}%
\hfill
\begin{minipage}[t]{0.48\columnwidth}
\centering
\begin{tabular}{ccc}
& TD & TS \\ \cline{2-3} 
\multicolumn{1}{c|}{TD} & \multicolumn{1}{c|}{-2, -2} & \multicolumn{1}{c|}{5, 3} \\ \cline{2-3} 
\multicolumn{1}{c|}{TS} & \multicolumn{1}{c|}{3, 5} & \multicolumn{1}{c|}{4, 4} \\ \cline{2-3} 
\end{tabular}
\end{minipage}
\caption{(Left) Chicken with Concede and (Right) a payments token game SPI on it}
\label{game:chicken_w_concede_and_payments_token_spi}
\end{table}

\paragraph*{Contributions}
In this work, we extend the token game framework for achieving SPIs in three ways, allowing players to additionally commit to (1) burn utility, (2) transfer utility to each other, and (3) transfer utility to outside parties. For each extension, we characterize the existence of SPIs and give computational complexity results.

\textbf{Money burning (\Cref{Sec:mb-token-SPIs}).} 
    We show that, perhaps surprisingly, money burning strictly increases the power of token games. 
    We characterize the existence of token game SPIs with money burning in two-player games (\Cref{thm:mb_token_spi_charactarization}),%
    and show that a large class of natural objectives over token game SPIs with money burning can be optimized efficiently using linear programming (\Cref{thm:mb_token_SPI_LP}).

\textbf{Utility transfers (\Cref{Sec:payment-token-SPIs}).} We characterize the existence of payments token game SPIs in n-player games (\Cref{thm:payment_token_SPI_charactarization}). 
    Notably, payments token SPIs can exist even in games where no actions are eliminated and no simple SPI exists. 
    We also give a linear program for optimizing over payments token SPIs (\Cref{thm:payment_token_SPI_LP}).

\textbf{Outside bettors (\Cref{Sec:outside_bettors}).} We introduce a scheme in which the players construct a financial security tied to the token game's outcome and sell it to an outside party. We show that token SPIs with outside bettors exist under extremely mild assumptions---essentially whenever the game is not constant sum (\Cref{thm:outside_bettors_SPI_existence2}), and how under stronger assumptions, they allow the players to achieve the maximum social welfare in the original game while preserving the original game's strategic structure and distribution of payoffs (\Cref{thm:outside_bettors_SPI_EV}).

\textbf{Computational Experiments (\Cref{Sec:experimental_results}).} We present computational experiments comparing the effectiveness of these interventions on randomly generated games. We find that the extensions introduced in this paper, especially token games with payments, substantially increase the frequency and magnitude of SPIs relative to prior methods.

\subsection{Related Work}
\label{sec:related-work}

As discussed in the introduction, SPIs were first introduced by \citet{SPI-original} and further studied by \citet{Sauerberg2025_promises_EPV-SPIs,digiovanni2024safe,Oesterheld2025}.
We borrow some of the conceptual features of this prior work, such as the idea of reducing games by elimination of irrational actions and using isomorphisms between games. However, we offer a simplified theoretical setup. Roughly, we directly define when we consider one game to be an SPI on another game (under the assumptions that strictly dominated actions can be removed and isomorphic games are played isomorphically). In contrast, prior papers separately introduce the assumptions and the SPI concept. Thus, our \textit{definitions} of SPIs (\Cref{def:iso-spi,def:simple-spi}) are characterization results in prior papers \cite[e.g.][Lemma 3.1]{Sauerberg2025_promises_EPV-SPIs}.
Our theoretical results primarily build on token games \citep{Sauerberg2025_promises_EPV-SPIs}, while our experimental results also cover utility function SPIs \citep{SPI-original}.

A lot of prior game-theoretic work has studied the use of commitments to money burning \cite{moulin1976_money_burning,money-burning-2x2} and to utility transfers 
\cite[e.g.][]{sauerberg2024commitment-to-payments,k-implementation,internal_implementation}.
However, none of it uses such commitment to achieve SPIs or uses token games.

Our \Cref{Sec:outside_bettors} shares some thematic connection to a line of work, initiated by \citet{Cavallo2006_redistribution_VCG_payments}, on incentive-compatibly redistributing budget surplus generated by the well-known VCG mechanism.
In particular, our outside bettors idea (\Cref{Sec:outside_bettors}) is reminiscent of the idea of choosing a player or set of players who cannot receive the items but are entitled to the entire budget surplus \cite{faltings2004_budget,moulin2009almost_BB_VCG_residual_claimant}. However, our surplus is generated through a very different mechanism: token games rather than payments in an auction.

\section{Preliminaries}

\subsection{Basic game theory}

An \emph{$n$-player (normal-form) game $\game$} is a pair $(A, \u)$, where $A = \bigtimes_i A_i$ for some nonempty set of \emph{actions} $A_i$ for each player $i$, and $\u: A \rightarrow \R^n$ is a \emph{utility function}. 
The elements of $A$ are \textit{outcomes} or action profiles, and $u_i(a)$ is Player $i$'s utility for outcome $a$. 
We use $\Delta(A)$ to denote the set of distributions over $A$, also called correlated strategy profiles. We extend $\u$ to distributions over outcomes by taking the expectation and define $\u(S) = \{\u(s) : s \in S\}$ for any set $S$ of outcomes or distributions over outcomes.  
We index the set of players $i \in \{1, \dots, n\} = [n]$, and use $-i$ to denote the set of players other than $i$. 
The (utilitarian) \textit{social welfare} $\SW(a)$ of an outcome $a$ is the sum of the players' payoffs $\SW(a) = \sum_{i\in [n]} u_i(a)$. 
We let $\SW(\cdot)$ apply to distributions over outcomes and payoff vectors as well. 
We also let $\SW^{max}(S)$ denote the maximum social welfare in the set $S$, $\max_{s \in S} \SW(s)$, with $\SW^{min}(A)$ defined analogously.

A \textit{subgame} $G'$ of an $n$-player game $\game=(A, \u)$ is a game $G' = (A', \u')$ where $A' = \bigtimes_i A'_i$ for nonempty subsets $A_i' \subseteq A_i$ and where $\u'$ is $\u$ restricted to $A'$. 
We imagine that Player $i$ is the same entity (person, etc.) in all games and that their utility is comparable between games. 
For this reason, we usually consider a single utility function $\u$ which applies to outcomes of any game. 
This allows us to reason about (safe) Pareto improvements between games.

An outcome \textit{$\a'$ is a (weak) Pareto improvement on $\a$}, denoted $\u(\a')\pgeq \u(\a)$,  if for all $i$ we have that $u_i(\a')\geq u_i(\a)$.
Furthermore, $\a'$ is a \textit{strict} Pareto improvement on $\a$, or $\u(\a') \pge \u(\a)$, if additionally there is a player $i$ for which $u_i(\a')>u_i(\a)$. 
We say that an outcome $\a$ is \textit{Pareto optimal} within a set $S$ if there is no $s \in S$ with $s \pge a $.
We also apply all of these terms to the payoff vectors $v = \u(a)$ themselves.

Let $\game$ be a game and let $a_i,a_i' \in A_i$ be actions for Player $i$. We say that \emph{$a_i'$ strictly dominates $a_i$} if for all $a_{-i}\in A_{-i}$ we have that $u_i(a_i',a_{-i}) > u_i(a_i,a_{-i})$.%

 A \textit{(game) isomorphism} from $(A,\mathbf u)$ to $(A',\mathbf u ')$ is a function $\iso\colon A \rightarrow A'$ defined by bijections $\iso_i: A_i \rightarrow A_i'$ such that there exist some $m, b \in \R^n$ with all $m_i > 0$ with $u_i'(\iso_1(a_1), %
..., \iso_n(a_n) %
) = m_i u_i(a) + b_i$ for all outcomes $a \in A$ and players $i \in [n]$.
An isomorphism is (weakly) Pareto improving if $u'_i(\iso(a)) \geq u_i(a)$ for all players $i$ and all $a \in A$ and strictly so if this inequality is strict for at least one player and outcome%
.
An isomorphism induces a function on utilities, parameterized by $m_i$ and $b_i$, which we call a utility correspondence.

\subsection{Safe Pareto improvements}

We now formally introduce SPIs.  Relative to prior work we have simplified the theoretical foundation (see \Cref{sec:related-work}).

Strategic interactions (as represented by normal-form games) can have actions that are clearly irrational, e.g., because they are strictly dominated by another action. One of the main ingredients to the SPI framework is that we are given some apparatus for removing such irrational actions.

\begin{definition}
    A \textit{reduction function} $\redf$ is a function that maps any game $G$ onto a subgame of $G$ s.t.\
    \begin{enumerate*}[(1)]
        \item for all games $G$ we have that $\redf(\redf(G))=\redf (G)$; and
        \item for all games $G$ and $G'$ with isomorphism $\phi$ we have that $\phi$ (restricted to the outcome set of $\redf (G)$) is also an isomorphism between $\redf (G)$ and $\redf (G')$.
    \end{enumerate*}
\end{definition}

Various ways to reduce games have been discussed in the literature. Unless otherwise specified, $\redf$ fully reduces the game by iterated elimination of strictly dominated strategies \cite{Pearce1984,Kohlberg1986}, which is the notion of reduction used by all prior SPI work. We will also consider the elimination of dominated strategies in \textit{mixed} strategies \cite{Pearce1984,Gilboa1990,Apt2004}. %
We'll often denote $\red G = \redf(G)$ and write $\red A$ for the action space of $\red G$.
Combining the notions of reduction and game isomorphism, 
we can now define isomorphism SPIs.
    
\begin{definition}\label{def:iso-spi}
    A game \emph{$G'$ is an isomorphism SPI on $G$} if $\redf(G')$ is isomorphic to $\redf(G)$ and this isomorphism is outcome-wise strictly Pareto improving.
\end{definition}

For detailed theoretical justifications of SPIs, see \citet{SPI-original,Oesterheld2025,Sauerberg2025_promises_EPV-SPIs}.
Note that there may be multiple isomorphisms between two given games. In \Cref{def:iso-spi} we only talk about \textit{one} of the isomorphisms. This is justified by the fact that all isomorphisms between two games must have the same effects on utilities, see \Cref{appendix_sec:multiple_isomorphisms}.
In addition to isomorphism SPIs, we consider simple SPIs, which for token games are essentially equivalent to agreeing to a single payoff profile. %

\begin{definition}\label{def:simple-spi}
    A game $G'$ is a \textit{simple SPI on $G$} if all outcomes in $\redf(G')$ are weakly Pareto better than all outcomes in $\redf(G)$ and this relation is strict for at least one pair. 
\end{definition}

\section{Token Game SPIs with Money Burning}\label{Sec:mb-token-SPIs}

In this section, we consider token games which are realized via joint commitments to play correlated strategy profiles in the base game and to burn utility as a function of the token outcome. 
Formally, given a base game $G = (A, \u)$, a \emph{token game with money burning} (or \emph{MB token game}) on $G$ is a game $\tgame = (T, \u)$ for which there exists  $\rf: T \rightarrow \C(A)$ such that $u_i(t) \leq u_i(\rf(t))$ for all players $i$ and token outcomes $t$. 
The players can realize such a token game by committing that, whenever outcome $t$ obtains, they'll play strategy profile $\rf(t)$ and each player $i$ will burn utility $u_i(\rf(t)) - u_i(t) \geq 0$, leaving each with the intended utility of $u_i(t)$.
\citeauthor{Sauerberg2025_promises_EPV-SPIs}'s (\citeyear{Sauerberg2025_promises_EPV-SPIs}) token games instead require that $u_i(t) = u_i(\rf(t))$. 

\begin{figure}[t]
\centering
\begin{tabular}{c|c|c|c|c|}
\multicolumn{1}{c}{} & \multicolumn{1}{c}{Hawk} & \multicolumn{1}{c}{Dove} & \multicolumn{1}{c}{Crazy} & \multicolumn{1}{c}{Refrain} \\
\cline{2-5}
Hawk & 0, 0 & 6, 2 & 0, 4 & 0, 0 \\
\cline{2-5}
Dove & 2, 6 & 4, 4 & 4, 0 & 0, 0 \\
\cline{2-5}
Crazy & 4, 0 & 0, 4 & 0, 0 & 0, 0 \\
\cline{2-5}
Superdove & $-2$, 12 & $-2$, 12 & $-2$, 12 & 5, 5 \\
\cline{2-5}
\end{tabular}

\includegraphics[scale=0.42]{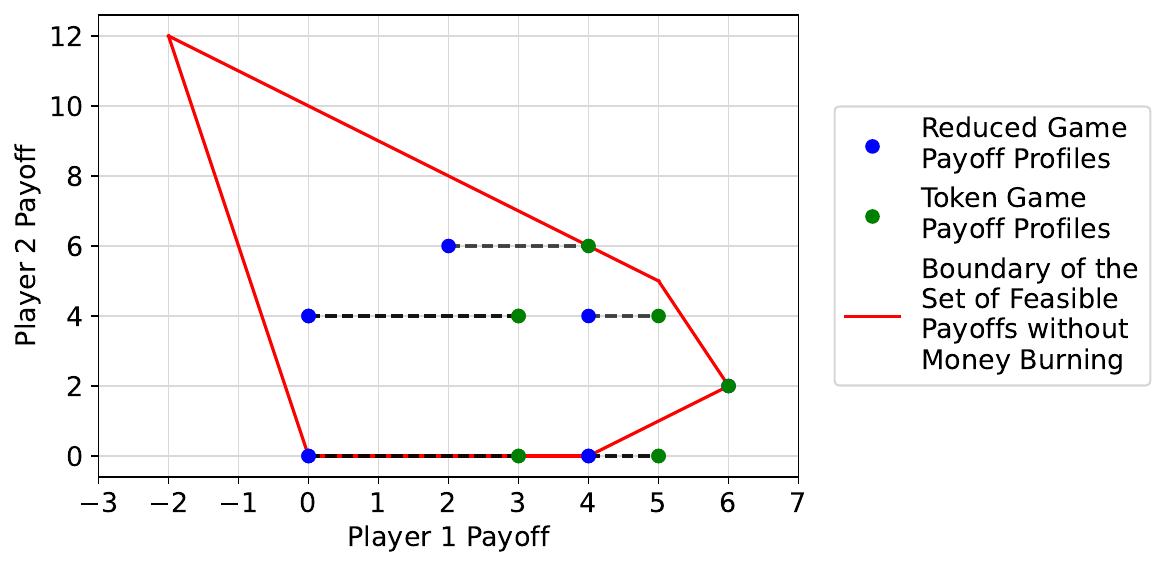}

\caption{The Hawk-Dove-Crazy Game and a geometric visualization of a token game SPI with money burning on it.}
\label{fig:HDC-combined}
\end{figure}

Perhaps surprisingly, commitment to burning money can enable new token SPIs.
As an example, consider the Hawk-Dove-Crazy game in \Cref{fig:HDC-combined}. The reduced game is a symmetric $3\times3$ game where each player has actions Hawk, Dove, and Crazy. %
It works roughly like rock/paper/scissors: Hawk beats Dove, Dove beats Crazy, and Crazy beats Hawk, though the exact payoffs vary. Roughly, if one keeps who \enquote{wins} the RPS aspect fixed, it's generally better for both if Dove is played and Crazy is not. 
In addition, Player $1$ has a \enquote{Superdove} action which gets exploited for $(-2, 12)$ by any of Player $2$'s normal actions but which results in payoffs of $(5,5)$ if Player $2$ refrains from exploiting. 
Refrain is strictly dominated for Player $2$ and therefore both Refrain and Superdove are eliminated under iterated strict dominance. 

We construct an MB token game isomorphic to the reduced Hawk-Dove-Crazy game. 
The isomorphism maps Player $2$'s payoffs by the identity and Player $1$'s payoffs by $v_1 \mapsto .5v_1 + 3$, mapping each halfway to P1's maximum payoff of $6$. 
This is visualized geometrically in \Cref{fig:HDC-combined}.

The $(4,6)$ payoff of (T Dove, T Hawk) can be realized by mixing between (Superdove, Refrain) and (Superdove, Hawk) in a 6:1 ratio for payoffs of $(6/7)(5, 5) + (1/7) (-2, 12) = (4,6)$.
The $(6,2)$ payoff of (T Hawk, T Dove) can be realized by playing (Hawk, Dove). 
All other token payoffs are Pareto dominated by the $(5,5)$ of (Superdove, Refrain) and so can be realized by commitments to play (Superdove, Refrain) and burn utility. 
Hence, this is a valid MB token game SPI. 

Interestingly, the ability to commit to burn utility is necessary to implement this, or any, token SPI on Hawk-Dove-Crazy. 
Geometrically, the key is the bottom right blue point at $(4,0)$, which this and any MB token SPI can and must improve beyond the non-money-burning feasible region. 
Below we give a relatively informal argument that money burning can be necessary for achieving token game SPIs; the formal proof follows from comparing \Cref{thm:mb_token_spi_charactarization} and Theorem 4.3 of \citet{Sauerberg2025_promises_EPV-SPIs}.

Consider the functions on the players' payoffs induced by an isomorphism. 
The $(6,2)$ payoff of (Hawk, Dove) is Pareto optimal in $G$, and hence must be a fixed point of each player's isomorphism. 
Hence, any SPI must improve only P1's payoffs while keeping P2's constant, and each of P1's other payoffs must be strictly improved. 
However, the $(4,0)$ payoff cannot be improved for Player $1$ without also increasing Player $2$'s payoff, and so Player $2$'s ability to commit to money burning is necessary for an SPI to exist.

We now turn to the questions of which games admit MB token game SPIs and of optimizing over such SPIs.  
It is easy to see that simple SPIs exist if and only if there is $\delta \in \Delta(A)$ which is Pareto better than all of $\red A$ and strictly so for at least one $a \in \red A$, so money burning makes no difference. 
Hence, we focus on isomorphism SPIs for the rest of this section.

Let $\tgame$ be an isomorphism SPI on $G$. Then by definition there exists an isomorphism $\phi$ between $\redf(G)$ and $\redf(\tgame)$.
We will associate with $\tgame$ the transformation on \textit{payoffs} induced by $\phi$.
This function, which we'll call a utility correspondence and denote $\remapu$, has two important properties. 
First, each payoff $\remapu(\u(a))$ for $a \in \red A$ must be attainable via commitments to play, at the corresponding token outcome $t$, the strategy profile $\remap(t)$ and burn the (nonnegative) utility vector $\u(\remap(t)) - \remapu(\u(a))$. 
In other words, $\remapu$ must be into the set of feasible payoffs $\fmb(G) = \{ v : \exists v' \in \u(\C(A)), v' \pgeq v\}$.
Secondly, by the definition of isomorphism, $\remapu$ must be player-wise positive affine, i.e. be of the form $\remapu_i (v) = m_i v_i + b_i$ for some  $m_i, b_i \in \R$ with $m_i > 0$ for each player $i$.
In other words, $\remapu$ must be a player-wise positive affine function $\remapu: \u(\red A) \rightarrow \fmb(G)$. 
We'll call such $\remapu$ ``valid''.
Of course, for any valid $\remapu$, we can construct an MB token game SPI which induces this utility correspondence. 
Therefore, when characterizing the existence of and optimizing over MB token game SPIs, we'll often just reason about the space of valid $\remapu$. 

Now, we introduce our characterization of MB token game SPIs in two-player games. 

\begin{theoremrep}\label{thm:mb_token_spi_charactarization}
    A two-player game $G$ admits a token game isomorphism SPI with money burning if and only if either (a) $\red{A}$ contains no outcomes which are Pareto optimal in $\F^{MB}(G)$ or (b) it contains exactly one payoff $v^*$ which is Pareto optimal in $\F^{MB}(G)$, this $v^*$ is at least one player's maximum payoff in $\red A$, and that player has at least one more distinct payoff in $\red G$.
\end{theoremrep}

\begin{proof}
    Let $V = \u(\red A)$ be the set of payoffs in the reduced game $\red G$. 
    Define $v_i^{max(G)} = \max_{v \in \u(A)} v_i$ to be each player's maximum payoff in $G$, and likewise $v_i^{max(V)} = \max_{v \in V} v_i$ to be each player's maximum payoff in the reduced game $\red G$. 
    We must decide whether there exists a valid $\remapu$: one which is player-wise positive affine and from $V$ into the feasible region $\fmb(G)$. 

    A valid (positive affine) $\remapu_i$ is determined by the choice of $\remapu_i(v_i)$ for two inputs $v_i$: there is only one line through those two points. 
    Much of the proof operates by reasoning about fixed points of $\remapu$. 
    For instance, if $v$ is Pareto optimal in $\fmb(G)$, it must be a fixed point of all $\remapu_i$: it cannot be increased in any dimension without being decreased in another dimension, which would make $\remapu$ fail to be everywhere improving for that player.     
    In addition, a valid $\remapu_i$ with $m_i \neq 1$ has exactly one fixed point, say $\check v_i$. 
    For an $\remapu_i$ which doesn't decrease any $v_i$ in $V$, this must either be mapping towards a high value, e.g. $v_i \mapsto v_i + \eps (\check v_i - v_i)$, or away from a low value, e.g. $v_i \mapsto v_i + \eps (v_i - \check v_i)$.
    
    \textit{If:}
    Consider the first case (a). Suppose $V$ contains no points which are Pareto optimal in $\fmb(G)$. 
    (Note that Pareto optimality in $\u(A)$ and $\fmb(G)$ are identical.)
    Then for all $v \in \u(\red A)$ for which $v_i \neq v_i^{max(G)}$ for either $i$, the feasible region contains an $\eps_v$ ball around $v$. 
    And any $v$ with $v_1 = v_1^{max(G)}$ must have $v_2 \neq v_2^{max}$ (or else $v$ would be Pareto optimal), and $v + (0, \eps_v)$ is feasible for some sufficiently small $\eps_v$.
    Of course, the case for $v_2 = v_2^{max(G)}$ is symmetric. 
    Therefore, $\remapu(v) = v + \eps (v^{max}-v)$ is feasible for some sufficiently small $\eps$.
    This is clearly strictly Pareto improving and entrywise positive affine, as desired. 

    Consider the second case (b). Suppose there is a single outcome $v^* \in \red A$ which is Pareto optimal in $\fmb(G)$, and that (without loss of generality) $v_1^*$ is Player $1$'s maximum payoff in $\red G$. 
    We claim that the $\remapu$ defined by $\remapu_2 = id$ and $\remapu_1(v_1) = (1-\eps) v_1 + \eps v_1^*$ is feasible. 
    Note that this would be strictly Pareto improving since Player $1$ has multiple distinct payoffs in $\red G$.
    We just need to show that, for all other $v' \neq v^*$, $v' + (\eps, 0)$ is feasible for some sufficiently small $\eps$ (which can depend on $v'$). 
    Because $v'$ is not Pareto optimal in $G$, there is some $v'' \pge v'$ and strictly so in at least one dimension. 
    If $v''_1 > v'_1$, we're immediately done. 
    Consider the case where all $v''$ $\u(A)$ which Pareto dominate $v'$ have $v'_1 = v''_1$. 
    Then in particular, the $v''$ which is on the Pareto frontier of $G$ must be the point on the Pareto frontier with maximum value in the first dimension, i.e. $v''_1 = v^{max(G)}_1 = v'_1$. 
    In this case, the proposed $\remapu$ has $\remapu(v') = v'$, which is feasible for any $\eps$. 
    Hence, the proposed $\remapu$ is feasible for some $\eps >0$ and there's an SPI, as desired.

    \textit{Only If: }
    Observe that any $v^*$ which is Pareto optimal in $\fmb(G)$ must be a fixed point of $\remapu$ for both players: it cannot be increased in one dimension without being decreased in the other, which would make $\remapu$ fail to be everywhere improving for that second player. 
    Therefore, if $v^*$ is intermediate for a player $i$, $\remapu_i$ must be the identity, as it's the only positive affine function which is (weakly) improving for both $v_i < v_i^*$ and $v_i> v_i^*$. 

    First, we consider the case where there's exactly one payoff $v^*$ in $V$ which is Pareto optimal in $G$ and where $v^*_i \neq v_i^{max(V)}$ for either player. 
    
    We proceed by cases. We know that $v^*$ is either minimal or intermediate in $V$ for both players. 
    First, if $v^*$ is intermediate for both players, we know $\remapu$ must be the identity for both players by the above logic, so no SPI exists. 
    If $v^*$ is minimal in $V$ for both players while being Pareto optimal, it would be the only payoff in $V$, and therefore again no SPI exists. 

    The only remaining case is where $v^*$ is minimal for (without loss of generality) P1 and intermediate for P2. 
    Since it's intermediate for P2 (in $V$), there must be some $v' \in \red A$ with $v'_2 > v^*_2$. Since $v^*$ is Pareto optimal, this implies that $v_1' < v_1^*$. But this contradicts that $v^*$ is minimal for P1. 
    Hence, this case cannot occur and no SPI exists, as desired. 

    Finally, we consider the case where there are multiple distinct payoffs in $V$ which are Pareto optimal in $\fmb(G)$. 
    In this case, each one must be a fixed point of each $\remapu_i$. 
    This immediately implies that each $\remapu_i$ must be the identity and no SPI exists, as desired. 
\end{proof}
\begin{proofsketch}
    The proof works by reasoning about the constraints that points $v \in \u(\red A)$ put on $\remapu$, and in particular any fixed points of $\remapu$.
    In particular, any $v^* \in \u(\red A)$ which is Pareto optimal in $G$ must be a fixed point of $\remapu$ (in both dimensions): it cannot be increased in either dimension without being decreased in the other, which cannot happen because $\remapu$ must be Pareto improving. 

    \textit{Only If / Nonexistence:}
    If $\u(\red A)$ has two distinct payoffs which are Pareto optimal in $G$, then each $\remapu_i$ has two fixed points and must be the identity. 
    In addition, say some Pareto-optimal $v^*$ is \enquote{intermediate} in $\u_i(\red A)$, i.e. Player $i$ has both higher and lower payoffs than $v_i^*$ in $\u(\red A)$. 
    This also implies that $\remapu_i$ must be the identity, since this $v_i^*$ is an intermediate fixed point and so any positive affine $\remapu_i$ aside from the identity would either decrease Player $i$'s payoff at values above or below $v_i^*$. 
    These together imply that, for an SPI to exist, there can be at most one $v^* \in \u(\red A)$ which is Pareto optimal in $G$, and if it exists it must be at least one player's maximum or minimum payoff. 
    But a Pareto-optimal $v^*$ cannot represent a player's minimum utility unless it's also a player's maximum utility, so that case is covered in the positive result.

    \textit{If / Existence:}
    In the case where $\red G$ contains no payoffs which are Pareto optimal in $G$, 
    we show that $\remapu(v) = v + \eps(v^{max} -v)$ is feasible for some $\eps > 0$. 
    Let $v^{max} = \left(\max_{v \in \u(A)} v_1, \max_{v \in \u(A)} v_2 \right)$ be the (typically infeasible) profile of each player's maximum payoff. 
    Geometrically, this corresponds to mapping each payoff some tiny fraction of the way towards this maximum point. 
    This is clearly feasible for any point $v$ on the interior of $\fmb(G)$. The boundary points which might pose a problem are those $v$ with $v_i = v_i^{max}$ for some $i$, but $\remapu$ only changes such $v$ in the other dimension, which is feasible since the points are not Pareto optimal.

    In the second case above, where $\red G$ contains a Pareto-optimal payoff $v^*$ where (without loss of generality) $v_1^*$ is Player $1$'s maximum payoff in $\red A$, we show that letting $\remapu_1(v_1) = v_1 + \eps(v_1^* - v_1)$ and $\remapu_2 = id$ is feasible for some $\eps > 0$. 
    Geometrically, this corresponds to projecting each point an $\eps$ fraction of the way towards the line defined by $v_1 = v_1^*$. 
    This is feasible for some nonzero $\eps$ because all remaining points $v$ in $\u(\red A)$ are Pareto dominated, so either they're Pareto dominated strictly in dimension $1$ or $v_1 = v_1^{max(G)}$, in which case $v_1^{max} = v_1^*$ and $\remapu$ is feasible for any $\eps$.   
\end{proofsketch}

We have shown that money burning makes token games more powerful. The following result, however, restricts the additional power from money burning in two-player games. In particular, it shows that when \citet{Sauerberg2025_promises_EPV-SPIs}'s token games don't yield an SPI, an MB token game SPI can only ever benefit one of the players.

\begin{corollaryrep}\label{cor:mb-can-enable-token-SPIs}
    If a two-player game $G$ does not admit a plain token game SPI, it cannot admit a token game SPI with money burning which is strict for both players. 
\end{corollaryrep}
\begin{proof}
    Consider a game $G$ which admits an MB token game SPI but not a plain token game SPI.     
    If $\red G$ contained no outcomes which were Pareto optimal in $G$, $G$ would admit a plain token game SPI by the characterization of plain token SPIs in \citet[Theorem 4.3]{Sauerberg2025_promises_EPV-SPIs}. 
    Hence, $G$ must be as in the second condition in \Cref{thm:mb_token_spi_charactarization}: it contains exactly one payoff $v^*$ which is Pareto optimal in $G$ and $v^*$ is (wlog) Player $1$'s maximum payoff in $\red G$. 
    Again by Theorem 4.3 of \citet{Sauerberg2025_promises_EPV-SPIs}, for such a $G$ not to admit a plain token SPI, this $v^*$ must be an intermediate payoff for Player $2$. 
    But then $v^*_2$ is an intermediate fixed point of $\remapu_2$, which therefore must be the identity and the SPI cannot be strict for Player $2$. 
\end{proof}

Although our characterization only applies to two-player games, we now show that,
for any number of players, linear programming can decide the existence of MB token
SPIs and optimize the large, natural class of min-linear objectives over them.
The class includes each player's utility, either in the worst case over outcomes or
in expectation under given beliefs (distributions over token outcomes, possibly
differing across players), measured as a level or as a gain over the base game, and
aggregated across players by weighted sum or by taking the minimum.
Formally, min-linear objectives are given by the minimum of a polynomially sized set of
linear functions of the variables $\remapu_i(\u(a))$ and $\remapu_i(\u(a)) -
u_i(a)$ for players $i$ and outcomes $a \in \red A$: the players' token game
payoffs and gains over the base game. 

\begin{theoremrep}\label{thm:mb_token_SPI_LP}
    It can be decided in polynomial time via linear programming whether a game $G$ admits a token game isomorphism SPI with money burning. 
    Furthermore, min-linear objectives over such SPIs can be optimized efficiently.
\end{theoremrep}
\begin{proof}
    We construct a polynomially sized linear program which essentially directly encodes the problem of finding or optimizing over valid $\remapu$. 
    Our LP is adapted from and extremely similar to that of \citet{Sauerberg2025_promises_EPV-SPIs} for the case of token SPIs without money burning. 
    We simply modify the feasibility constraints to allow for money burning.
    
    Our LP has several types of variables: 
    For each player $i$, we have variables $m_i$ and $b_i$ which correspond to the parameters of the utility correspondence $\remapu_i$. 
    Our LP takes as input the payoff vectors in the reduced game $\red G$, which we denote $v^1, \dots, v^k$. 
    For each $v^j$, we have a set of variables $p_a^{v^j}$ for all $a \in A$ which collectively represent the strategy profile in $G$, i.e. distribution over outcomes in $A$, which proves the feasibility of a token payoff $\remapu(v^j)$. 
    The utility function $\u: A \rightarrow \R$ is also an input. 

\begin{align*}
    &\text{Maximize}\quad \sum_{v^j \in V} \sum_{i \in [n]} (m_i v_i^j + b_i - v_i^j) &\\
    &\text{Subject to:} &\\
    & m_i \geq 0 &\text{for all } i \in [n]\\
    & m_i v_i^j + b_i \geq v_i^j &\text{for all } i \in [n], j \in [k] \\
    & p_a^{v^j} \geq 0 &\text{for all } a \in A, j \in [k]  \\
    &\sum_{a \in A} p_a^{v^j} = 1 &\text{for all } j \in [k] \\
    & \sum_{a \in A} p_a^{v^j} u_i(a) \geq m_i v_i^j + b_i &\text{for all } i \in [n], j \in [k]
\end{align*}

    The first constraint ensures that the $\remapu$ is (weakly) positive affine. 
    The second constraint ensures that $\remapu$ is (weakly) Pareto improving on all $v^j$, which represent payoff profiles in $\red G$. 
    The next two constraints collectively ensure that for each $v^j$, the set $\{p_a^{v^j}\}$ corresponds to a valid probability distribution over outcomes in $G$, i.e. $\remap(\cdot)$. 
    And the final constraint ensures the feasibility of $\remapu(v^j)$ for each $j$: it requires that for each $v^j$ and each Player $i$, Player $i$'s expected utility for the distribution $\{p_a^{v^j}\}$ is at least that specified by $\remapu$ for the token outcome corresponding to $v^j$, and therefore the exact payoff specified by $\remapu$ can be realized by playing the distribution and committing to burn the difference between the utilities. 
    
    It's easy to verify that the program is linear (recall that the sets $A$ and vectors $v^j$ and $u_i(a)$ are inputs to the program) and polynomially sized: it has $O(|A|^2)$ variables and $O(|A|^2)$ constraints. 
    Therefore, the objective of the LP can be optimized in polynomial time.

    Now, we claim that a game $G$ admits an MB token game isomorphism SPI if and only if the optimal value of the LP is strictly greater than $0$. 
    The objective value of the LP is exactly the sum over players and payoffs of $\remapu_i(v^j_i) - v^j_i$, i.e. the social welfare gain on payoff vector $v^j$ from the SPI. 
    Since each of these is non-negative, the objective is strictly positive if and only if the $\remapu$ is strictly Pareto improving. 

    The only remaining subtlety is that our program requires only that $m_i \geq 0$, but valid utility correspondences must be strictly positive affine.
    This can be dealt with as in \citet{Sauerberg2025_promises_EPV-SPIs}.
    
    First, observe that the space of valid $\remapu$ is convex. 
    For any pair of valid $\remapu$, say $\remapu^1(v)= m^1 v + b^1$ and $\remapu^2(v)= m^2 v + b^2$, $\alpha \remapu^1 + (1-\alpha) \remapu^2$ is simply $(\alpha m^1 + (1-\alpha)m^2) v + \alpha b^1 + (1-\alpha) b^2$. 
    In particular, the convex combination still has each $m_i > 0$ and is positive affine. 
    The convex combination is also still feasible in $\fmb(G)$ because the underlying space $\fmb$ itself is convex. 

    Also, note that the identity utility correspondence is always feasible (though it is not a strict SPI).
    Therefore, if the LP finds a solution corresponding to a $\remapu$ with some $m_i =0$, we can construct a valid $\remapu'(v)$ by $\remapu'(v)= (1-\eps) \remapu(v) + \eps v$. 
    This $\remapu'$ has all $m_i \geq \eps$, and is still strictly Pareto improving if $\remapu$ is. 
    Hence, we've successfully reduced deciding the existence of an MB token game isomorphism SPI to checking the optimal objective value of the LP, which can be done in polynomial time, as desired. 

    \textit{Optimization:}
    Consider a game $G$ and a min-linear objective defined by a set of linear functions $F = \{f^1, \dots, f^{|F|}\}$, where each $f^k$ is a linear function of variables of the form $\remapu_i(\u(a))$ or $\remapu_i(\u(a)) - u_i(a)$ for $a \in \red A$ and $i \in [n]$.
    Let's write $w_i^k(v^j)$ to denote the sum of the coefficients on all $\remapu_i(\u(a))$ in $f^k$ for all $a$ with $\u(a) = v^j$. 
    (This is only necessary because our LP does not distinguish between $\remapu_i(\u(a))$ for different outcomes $a$ with the same payoff vector $\u(a)$. 
    However, all such outcomes have the same image under the utility correspondence, so this is not an issue.)
    Similarly, let's write $\tilde w_i^k(v^j)$ to denote the sum of the coefficients on all $\remapu_i(\u(a)) - u_i(a)$ in $f^k$ for all $a$ with $\u(a) = v^j$.

    We construct an LP which is essentially the same as the one above, except that we replace the objective with a new variable $\alpha$ and add constraints that $\alpha \leq f^k$ for each $k \in [|F|]$.

    \begin{align*}
    &\text{Maximize}\quad \alpha &\\
    &\text{Subject to:} &\\
    & m_i \geq 0 &\text{for all } i \in [n]\\
    & m_i v_i^j + b_i \geq v_i^j &\text{for all } i \in [n], j \in [k] \\
    & p_a^{v^j} \geq 0 &\text{for all } a \in A, j \in [k]  \\
    &\sum_{a \in A} p_a^{v^j} = 1 &\text{for all } j \in [k] \\
    & \sum_{a \in A} p_a^{v^j} u_i(a) \geq m_i v_i^j + b_i &\text{for all } i \in [n], j \in [k]\\
    & \alpha \leq \sum_{i \in [n]} \sum_{v^j \in V} w_i^k(v^j) (m_i v_i^j + b_i) \\ &\hspace{.75in} +\tilde w_i^k(v^j) (m_i v_i^j + b_i - v_i^j) &\text{for all } k \in [|F|]
\end{align*}

    Note that our LP is still linear: the $w_i^k(v^j)$ and $\tilde w_i^k(v^j)$ are inputs to the program, so the last set of constraints are linear in the variables $m_i$, $b_i$, and $\alpha$.
    Our program is also still polynomially sized, since $|F|$ is polynomial.

    Because the LP relaxes $m_i > 0$ and strict Pareto improvement to weak inequalities, its optimal solution need not be an SPI, and since the space of valid $\remapu$ is not closed, an optimal SPI need not exist.
    We handle this as \citet{Sauerberg2025_promises_EPV-SPIs} do.
    Assume $G$ admits an MB token game isomorphism SPI (decidable as above), since otherwise the optimization problem is infeasible.
    Add a variable $\mu$ with constraints $\mu \leq m_i$ for all $i \in [n]$ and $\mu \leq \sum_{v^j \in V}\sum_{i \in [n]} (m_i v_i^j + b_i - v_i^j)$, and maximize $(\alpha, \mu)$ lexicographically, e.g.\ by first computing the optimal value $\alpha^*$ and then maximizing $\mu$ subject to $\alpha \geq \alpha^*$.
    If the optimal $\mu$ is positive, the solution has all $m_i > 0$ and is strictly Pareto improving, so it is an optimal SPI.
    If the optimal $\mu$ is $0$, no SPI attains $\alpha^*$, since any SPI attaining it would yield a feasible solution with $\mu > 0$.
    In that case $\alpha^*$ is still the supremum over SPIs: let $\remapu^*$ be the LP solution, let $\remapu^s$ be any SPI with all $m_i > 0$ (found as above), and let $\remapu^\eps(v) = (1-2\eps)\remapu^*(v) + \eps\remapu^s(v) + \eps v$.
    $\remapu^\eps$ is feasible by convexity of the LP's feasible region, has all $m_i \geq \eps > 0$, and is strictly Pareto improving because $\remapu^s$ is and the other two terms are weakly Pareto improving.
    Its objective value tends to $\alpha^*$ as $\eps \to 0$.
\end{proof}

Optimizing over these SPIs involves a subtlety.
As with the token SPIs of \citet{Sauerberg2025_promises_EPV-SPIs}, an optimal SPI may not exist, since the space of SPIs is not closed: their utility correspondences must have $m_i > 0$ and be strictly Pareto improving.
Nonetheless, our LP approach efficiently decides whether an optimal SPI exists and finds one if so; otherwise, it computes the supremum of the objective and a family of SPIs whose values approach it.

\section{Token Game SPIs with Payments}\label{Sec:payment-token-SPIs}

In this section, we consider the token games which are realized by joint commitments to play correlated strategy profiles and transfer utility as a function of the token outcome. 
We assume players can transfer utility to each other losslessly, linearly, and without any budget constraints. 

Formally, given a base game $G = (A, \u)$, a \emph{payments token game} on $G$ is a game $\tgame = (T, \u)$ for which $\SW(t) \in \left[\min_{a \in A} \SW(a), \max_{a\in A} \SW(a)\right]$ for all $t \in T$.
(Recall that $\SW(a) = \sum_i u_i(a)$ denotes the utilitarian social welfare of an outcome.)
These token payoffs are exactly those which can be realized by a commitment to a correlated strategy profile and payment vector, i.e. the $\u(t)$ such that $\u(t) = \u(\rf(t)) + p(t)$ for some $\rf(t) \in \Delta(A)$ and $p(t) \in \R^n$ with $\sum_i p_i(t) =0$.

Note that additionally allowing commitments to money burning makes no difference here. Any payoff profile with social welfare within $\left[\min_{a \in A} \SW(a), \max_{a\in A} \SW(a)\right]$ can already be achieved with transfers alone.
Thus, the only additional payoff profiles made feasible by money burning are those with social welfare less than $\min_{a \in A} \SW(a)$, which are useless for token SPIs because they'd necessarily leave at least one player worse off than in the original game.

\begin{table}[t]
\centering
\begin{minipage}[t]{0.48\columnwidth}
\centering
\begin{tabular}{ccc}
& Dare                      & Swerve                    \\ \cline{2-3} 
\multicolumn{1}{c|}{Dare}   & \multicolumn{1}{c|}{-9, -9} & \multicolumn{1}{c|}{6, 1} \\ \cline{2-3} 
\multicolumn{1}{c|}{Swerve} & \multicolumn{1}{c|}{1, 5} & \multicolumn{1}{c|}{3, 3} \\ \cline{2-3} 
\end{tabular}
\end{minipage}%
\hfill
\begin{minipage}[t]{0.48\columnwidth}
\centering
\begin{tabular}{ccc}
& T Dare                      & T Swerve                    \\ \cline{2-3} 
\multicolumn{1}{c|}{T Dare}   & \multicolumn{1}{c|}{-6, -9} & \multicolumn{1}{c|}{6, 1} \\ \cline{2-3} 
\multicolumn{1}{c|}{T Swerve} & \multicolumn{1}{c|}{2, 5} & \multicolumn{1}{c|}{3.6, 3} \\ \cline{2-3} 
\end{tabular}
\end{minipage}
\caption{(Left) Asymmetric Chicken and (Right) a payments token game SPI on Asymmetric Chicken}
\label{game:asymmetric_chicken_and_p_token_spi}
\end{table}

For example, consider Asymmetric Chicken in \Cref{game:asymmetric_chicken_and_p_token_spi}. 
The only asymmetry is that Player $1$'s utility for (Dare, Swerve) is higher than Player $2$'s for (Swerve, Dare). 
(Dare, Swerve) is the unique social-welfare-maximizing outcome. 
The token game (Right) is a payments (isomorphism) token game SPI on Asymmetric Chicken. 
The utility correspondence is given by $\remapu_1(v_1) = 0.8 v_1 + 1.2$ and $\remapu_2(v_2) = v_2$, so the SPI is only strict for Player $1$. 
It's easy to verify that each token payoff has social welfare of at most $7$, the maximum social welfare in Asymmetric Chicken, and is hence feasible. 
Note also that this token SPI is pointwise optimal: no payments isomorphism token SPI on Asymmetric Chicken gives either player a higher token payoff at any outcome.
Since (Dare, Swerve) $= (6,1)$ maximizes social welfare, its token payoff must remain $(6,1)$, and since $1$ is an intermediate payoff for Player $2$, $\remapu_2$ must be the identity.
Feasibility at (Swerve, Dare) $= (1,5)$ then requires $\remapu_1(1) \leq 2$, i.e., $m_1 \geq 0.8$.
Since Player $1$'s other payoffs all lie below $6$, $m_1 = 0.8$ maximizes each of them.

Asymmetric Chicken admits an isomorphism payments token game SPI even though the reduction function doesn't eliminate any actions.
Such games do not admit non-simple SPIs under most previously considered forms of commitment, including plain token SPIs, MB token SPIs (\Cref{Sec:mb-token-SPIs}), and utility function SPIs \cite{SPI-original}.

We now turn to characterizing and computing payments token SPIs. 
Again there's a simple characterization of \textit{simple} token game SPIs: they exist essentially whenever a single payoff profile can simultaneously give each player their maximum payoff in $\red G$. 
The only subtlety is ensuring strictness.

\begin{lemmarep}\label{simple_payemnts_token_SPI}
     A game $G$ admits a simple  payments token SPI if and only if either its reduced game has multiple distinct payoff profiles and $\sum_i \max_{a \in \red A} u_i(a) \leq \SW^{max}(G)$ or if its reduced game has exactly one distinct payoff profile and $\sum_i \max_{a \in \red A} u_i(a) < \SW^{max}(G)$.
\end{lemmarep}
\begin{proof}
    \textit{If:} 
    First, in the case where $\red G$ has exactly one distinct payoff profile, suppose $\sum_i \max_{a \in \red A} u_i(a) < \sum_i \max_{a \in \red A} u_i(a) + c = \SW^{max}(G) $ for some constant $c$. Then the token game which consists of a single outcome $t$ and where $\u_i(t) = \max_{a \in \red A} u_i(a) +c/n$ is a simple payments token game SPI on $G$. 

    Now, consider the case where $\red G$ has multiple distinct payoff profiles, and $\sum_i \max_{a \in \red A} u_i(a) = \SW^{max}(G)$.
    The token game which consists of a single outcome $t$ and where $\u_i(t) = \max_{a \in \red A} u_i(a)$ is a simple payments token game SPI on $G$. 
    This outcome $t$ is clearly weakly Pareto improving on all $a \in \red A$. 
    And the outcome is strictly Pareto improving on any $a \in \red A$ in which at least one player fails to achieve their maximum utility in $\red A$, which must exist because $\red A$ contains multiple distinct payoff profiles and hence at least one player who sometimes receives less than their maximum utility. 

    \textit{Only If:} 
    Suppose $\tgame = (T, \u)$ is a simple payments token SPI on $G$. 
    Then by definition, there is a token outcome $t$ such that $\u_i(t) \geq u_i(a)$ for all $a \in \red A$,
    and this inequality is strict for at least one outcome $a^* \in \red A$ and player $i$. 

    The first condition implies that $\u_i(t) \geq \max_{a \in \red A} u_i(a)$ for all $i$. Summing both sides of the inequality over all $i \in [n]$ and noting that $\SW^{max}(G) \geq \SW(\u(t))$ yields that 
    $\SW^{max}(G) \geq \sum_i \max_{a \in \red A} u_i(a)$, as desired. 
    For the case where $\red G$ has exactly one distinct payoff profile, the second condition says that $\u_i(t) > \max_{a \in \red A} u_i(a)$ for some $i$, and therefore summing both sides of the inequality above over $i$ yields a strict inequality. 
    We conclude that $\SW^{max}(G) \geq \SW(\u(t)) > \sum_i \max_{a \in \red A} u_i(a)$, as desired. 
\end{proof}

Therefore, we focus on isomorphism SPIs for the rest of the section. 
As in \Cref{Sec:mb-token-SPIs}, we associate each isomorphism SPI with its utility correspondence $\remapu$, which captures the isomorphism's effect on the players' base game payoffs. 
By the definition of isomorphism, $\remapu$ is required to be player-wise positive affine, and is required to be Pareto improving on $\u(\red A)$ (weakly everywhere and strictly on at least one point) by the definition of SPI. 
Finally, $\remapu$ can be achieved by a payments token game if and only if $\SW(\remapu(v)) \leq \SW^{max}(G)$ for all $v \in \u(\red A)$. 
We call such a utility correspondence valid and reason about payments token game SPIs by reasoning about the space of valid $\remapu$.

\begin{theoremrep}\label{thm:payment_token_SPI_charactarization}
    A game $G$ admits a payments token SPI if and only if either no payoff profiles in $\u(\red A)$ are social welfare maximizing in $\u(A)$ or all three of the following hold: 
    \begin{enumerate*}[(1)]
        \item some Player $i$ achieves the same payoff $v_i^*$ in all social-welfare-maximizing payoff profiles in $\u(\red A)$, 
        \item this $v_{i}^*$ is either Player $i$'s maximum or minimum payoff in $\u(\red A)$, and 
        \item Player $i$ has at least one other payoff in $\u(\red A)$.
    \end{enumerate*}
\end{theoremrep}

\begin{proof} 
Let $V = \u(\red A)$ and let $V^*$ be the subset of $V$ which are social welfare maximizing in $\u(A)$. 

\textit{If:}
First, consider the case where $V^* = \emptyset$. 
Let $d_v = \SW^{max}(G) - \SW(v)$ for each $v \in V$, and let $d = \min_{v \in V} d_v$. 
Then $\remapu(v) = v + (d/n, d/n, \dots, d/n)$ is a valid utility correspondence, and hence $G$ admits a payments token SPI, as desired. 

Now, consider the second case. 
Suppose $|V^*| \geq 1$, and that (without loss of generality) Player $1$ has the same payoff $v^*_1$ in all payoff profiles in $V^*$. 
For any $v$ with $v_1 \neq v_1^*$, $\SW(v) < \SW^{max}(G)$. 
Let's say $\SW(v) = \SW^{max}(G) - d_v$.
We claim that there's a valid $\remapu$ where $\remapu_i = id$ for all $i \neq 1$, $v_1^*$ is a fixed point of $\remapu_i$, and $\remapu_1(v_1) > v_1$ for all other $v_1 \neq v_1^*$. 
In particular, if $v_1^* = \max_{v' \in V} v_1'$, then the $\remapu_1(v_1)$ defined by $\remapu_1(v_1) = v_1 + \eps (v_1^* -v_1)$ is feasible where $\eps = \min_v \frac{d_v}{v_1^* - v_1} > 0$ since this implies $\remapu_1(v_1) \leq v_1 + d_v$ for all $v \in V$ and therefore $\SW(\remapu(v)) \leq \SW(v) + d_v \leq \SW^{max}(G)$.
Similarly, if $v_1^* = \min_{v' \in V} v_1'$, then the $\remapu_1(v_1)$ defined by $\remapu_1(v_1) = v_1 + \eps (v_1 - v_1^*)$ is feasible for $\eps = \min_v \frac{d_v}{v_1 - v_1^*} > 0$.
In either case, we have a feasible, player-wise positive affine $\remapu$ which is strictly Pareto improving because Player $1$ has at least one payoff aside from $v_1^*$ in $\u(\red A)$.
Hence, we have a payments token SPI, as desired. 

\textit{Only If: }
Suppose neither of the above conditions apply. 
That is, for all Players $i$, either (1) Player $i$ achieves multiple distinct payoffs in $V^*$, (2) Player $i$ achieves a payoff in $V^*$ which is neither minimal nor maximal in $u_i(\red A)$, or (3) Player $i$ has one payoff in $V^*$ and this is their only payoff in $\u(\red A)$. 
In each case, we show that the only valid $\remapu_i$ is the identity, and therefore $G$ admits no strict payments SPI. 

As usual, each point in $V^*$ must be a fixed point of $\remapu$ and hence $\remapu_i$.
In case (1), each of Player $i$'s payoffs in $V^*$ must be a fixed point of $\remapu_i$ and the only positive affine $\remapu_i$ with multiple fixed points is the identity.    
In case (2), $v_i^*$ is a fixed point of $\remapu_i$ and there are $v_i \in V$ both strictly greater than and less than $v_i^*$. The only positive affine $\remapu_i$ with $v^*_i$ as a fixed point and which doesn't have $\remapu_i(v_i) < v_i$ on either side of $v^*_i$ is the identity. 
Finally, in case (3), $v^*_i$ must be a fixed point of $\remapu_i$ and Player $i$ has no other payoffs in $\u(\red A)$, so again $\remapu_i$ is the identity on $\u(\red A)$. 
\end{proof}

\begin{proofsketch}
    \textit{Only if / Nonexistence:}
    The nonexistence results rely on the fact that any payoff profile $v^*$ which is social welfare maximizing in $A$ must be a fixed point of $\remapu$ (in all dimensions), since if $\remapu(v^*)$ is greater than $v^*$ for one player, it would necessarily be less than $v^*$ for another player and fail to be Pareto improving. 
    We show that for any Player $i$ who fails to meet any of the three conditions above, $\remapu_i$ must be the identity.
    If (1) fails, $\remapu_i$ must be linear (positive affine) and have at least two distinct fixed points.  
    If (2) fails, $\remapu_i$ must be linear (positive affine) with a fixed point at $v_i^*$ while having $\remapu_i(v_i) \geq v_i$ both above and below $v_i^*$. 
    If (3) fails, then $v_i^*$ is a fixed point of $\remapu_i$ and also the only input. 

    \textit{If / Existence:}
    If no payoff profiles in $\u(\red A)$ are social welfare maximizing in $\u(A)$, then $\remapu(v) = v + (\eps, \eps, \dots, \eps)$ is feasible for some sufficiently small $\eps$. 
    In the other case, we construct a valid $\remapu$ which is the identity for all players but Player $i$, where $\remapu_i$ has a fixed point at $v_i^*$ and has $\remapu_i(v_i) > v_i$ for all other $v_i \in \u_i(\red A)$. 
    In the case where $v_i^*$ is Player $i$'s maximum payoff in $\u(\red A)$ we achieve this by $\remapu_i(v_i) = v_i + \eps (v_i^* -v_i)$ for some small $\eps >0$,and when it's Player $i$'s minimum payoff, by $\remapu_i(v_i) = v_i + \eps (v_i - v_i^*)$. 
    Geometrically, these can be thought of as mapping each $v_i$ either an $\eps$-fraction of the way towards a maximum or an $\eps$-fraction of the way away from a minimum. 
\end{proofsketch}

As with the MB token SPIs, we can optimize min-linear objectives over payments token SPIs via linear programming.
\begin{theoremrep}\label{thm:payment_token_SPI_LP}
    It can be decided in polynomial time via linear programming whether a game $G$ admits a payments token game isomorphism SPI. 
    Furthermore, min-linear objectives over such SPIs can be optimized efficiently.
\end{theoremrep}
\begin{proof}

We construct a linear program which directly encodes the problem of finding or optimizing over valid $\remapu$. 
Our LP is substantially simpler than that of \Cref{thm:mb_token_SPI_LP} because feasibility of token payoffs under payments is determined entirely by a social welfare bound, so no auxiliary distribution variables are needed.

Its only variables are $m_i$ and $b_i$ corresponding to the parameters of each $\remapu_i(v_i) = m_i v_i + b_i$ for each player $i$. 
The payoff profiles $v^j$, which we index $v^1, \dots, v^k$, are inputs to the LP and constitute the set of payoff profiles in $\u(\red A)$.  

\begin{align*}
    &\text{Maximize}\quad \sum_{v^j \in V} \sum_{i \in [n]} (m_i v_i^j + b_i - v_i^j) &\\
    &\text{Subject to:} &\\
    & m_i \geq 0 &\text{for all } i \in [n]\\
    & m_i v_i^j + b_i \geq v_i^j &\text{for all } i \in [n], j \in [k] \\
    & \sum_{i \in [n]} (m_i v_i^j + b_i) \leq \SW^{max}(G) &\text{for all } j \in [k]
\end{align*}

It's easy to see that the program is linear; recall that the payoff profiles $v^j$ are parameters of the problem / inputs to the program. 
The LP has $O(n)$ variables and $O(n \cdot k)$ constraints, both of which are polynomial in the instance size, so the problem can be optimized in polynomial time. 

The LP essentially directly encodes the constraints and optimizes over $\remapu$.
The first set of constraints ensures that each $\remapu_i$ is (weakly) positive affine. 
The second set of constraints ensures that $\remapu$ is weakly Pareto improving on each $v^j$. 
The final set of constraints ensures that the total social welfare of the token payoff corresponding to each $v^j$ does not exceed $\SW^{max}(G)$, which is necessary and sufficient for feasibility with payments.
This suffices for feasibility because the constraint that the social welfare of each token payoff is at least the minimum social welfare in $G$ is already implied by the constraint that $\remapu$ be weakly Pareto improving on each $v^j$. 

Now, we claim that a game $G$ admits a payments token game isomorphism SPI if and only if the optimal value of the LP is strictly greater than $0$.
The objective value of the LP is exactly the sum over players and payoff profiles of $\remapu_i(v^j_i) - v^j_i$, i.e.\ the total social welfare gain across all payoff profiles from the SPI.
Since each term is non-negative (by the Pareto-improving constraints), the objective is strictly positive if and only if the $\remapu$ is strictly Pareto improving.

The only remaining subtlety is that our program requires only that $m_i \geq 0$, but valid utility correspondences must be strictly positive affine.
As in the proof of \Cref{thm:mb_token_SPI_LP}, the space of valid $\remapu$ is convex, and the identity utility correspondence is always feasible (though it is not a strict SPI).
Therefore, if the LP finds a solution corresponding to a $\remapu$ with some $m_i = 0$, we can construct a valid $\remapu'(v) = (1 - \eps) \remapu(v) + \eps v$ for small $\eps > 0$.
This $\remapu'$ has all $m_i \geq \eps$ and is still strictly Pareto improving if $\remapu$ is.
Moreover, the convex combination preserves feasibility because if $\sum_i \remapu_i(v^j_i) \leq \SW^{max}(G)$ and $\sum_i v^j_i \leq \SW^{max}(G)$, then $\sum_i \remapu'_i(v^j_i) \leq \SW^{max}(G)$.
Hence, we've successfully reduced deciding the existence of a payments token game isomorphism SPI to checking the optimal objective value of the LP, which can be done in polynomial time, as desired.

    \textit{Optimization:}
    Consider a game $G$ and a min-linear objective defined by a set of linear functions $F = \{f^1, \dots, f^{|F|}\}$, where each $f^\ell$ is a linear function of variables of the form $\remapu_i(v^j_i)$ or $\remapu_i(v^j_i) - v^j_i$ for $v^j \in \u(\red A)$ and $i \in [n]$.
    Let $w_i^\ell(v^j)$ denote the coefficient on $\remapu_i(v^j_i)$ in $f^\ell$, and let $\tilde w_i^\ell(v^j)$ denote the coefficient on $\remapu_i(v^j_i) - v^j_i$ in $f^\ell$.

    We construct an LP with the same feasibility constraints as above, but replace the objective with a new variable $\alpha$ and add constraints that $\alpha \leq f^\ell$ for each $\ell \in [|F|]$:

    \begin{align*}
    &\text{Maximize}\quad \alpha &\\
    &\text{Subject to:} &\\
    & m_i \geq 0 &\text{for all } i \in [n]\\
    & m_i v_i^j + b_i \geq v_i^j &\text{for all } i \in [n], j \in [k] \\
    & \sum_{i \in [n]} (m_i v_i^j + b_i) \leq \SW^{max}(G) &\text{for all } j \in [k] \\
    & \alpha \leq \sum_{i \in [n]} \sum_{v^j \in V} w_i^\ell(v^j) (m_i v_i^j + b_i) \\ &\hspace{.75in} + \tilde w_i^\ell(v^j) (m_i v_i^j + b_i - v_i^j) &\text{for all } \ell \in [|F|]
\end{align*}

    Note that our LP is still linear: the $w_i^\ell(v^j)$ and $\tilde w_i^\ell(v^j)$ are inputs to the program, so the last set of constraints are linear in the variables $m_i$, $b_i$, and $\alpha$.
    Our program is also still polynomially sized, since $|F|$ is polynomial.
    We handle optimality exactly as in the proof of \Cref{thm:mb_token_SPI_LP}: that argument uses only that the LP's feasible region is convex, that the identity is feasible, and that an SPI with all $m_i > 0$ can be found by the existence LP, all of which hold here.
\end{proof}

\section{Token Game SPIs with Outside Bettors}\label{Sec:outside_bettors}

One can view the payoffs in the payments token games from the previous section as each realized via a strategy profile $\rf(t)$ and a net payment vector $p(t)$ which sums to $0$.  
Because the token payoffs $\u(t)$ are not all social welfare maximizing, these strategy profiles $\rf(t)$ are also not always social welfare maximizing.
The players could instead realize the token payoffs by committing to play some social-welfare-maximizing profile, make some payments to each other to realize the token payoffs, and then ``give away" the excess utility by making payments to parties outside the game. 
However, there's an apparent problem: the players cannot give it to anything they value without violating the isomorphism constraint, for the same reason they cannot keep it for themselves. 

We now develop a way for the players to benefit from this excess utility by
extending the transferable utility assumption from the previous section:
Utility can now be transferred between the players in the game and an outside
party, whom we call an \emph{outside bettor}.
In particular, we imagine the players construct and sell a security (i.e., a financial
instrument) whose value depends on the outcome of a token game.

An outside bettors token game on $G$ is constructed as follows.
The players create a security and, for each outcome $a$ of $G$, commit that at
the corresponding token outcome they will play a fixed social-welfare-maximizing
outcome $a^*$, pay the holder of the security $\SW(a^*) - \SW(a)$, and make
payments so that each player's net payoff is $u_i(a)$.
Note that the security never pays out a negative amount, since $a^*$ is social
welfare maximizing.
They then sell the security to an outside bettor for a price $x \geq 0$, split
the proceeds equally so each player gains $x/n$ regardless of the token
outcome, and play the token game as usual.
Including the proceeds, the result is a token game with payoffs
$\u(a) + (x/n, \dots, x/n)$ and utility correspondence
$\remapu(v) = v + (x/n, \dots, x/n)$.
Such an outside bettors token game is an (isomorphism) SPI on $G$ whenever $x>0$.
This claim requires that selling the security not influence how the token game is played, which we ensure with three assumptions.
First, other than committing to the token game itself, the players make no commitments before the security is sold. Otherwise, they would have an incentive to raise the sale price $x$ by committing to strategies that make low-welfare outcomes more likely.
Second, the bettor cannot influence play: buying the security is their only interaction with the players.
Third, the players do not update their beliefs about each other's strategies based on $x$. This can be ensured by holding the proceeds in escrow, so that $x$ stays hidden until after play.

\begin{toappendix}
\begin{table}[t]
\centering
\begin{tabular}{ccc}
& Dare                      & Swerve                    \\ \cline{2-3} 
\multicolumn{1}{c|}{Dare}   & \multicolumn{1}{c|}{-9, -9} & \multicolumn{1}{c|}{5, 1} \\ \cline{2-3} 
\multicolumn{1}{c|}{Swerve} & \multicolumn{1}{c|}{1, 5} & \multicolumn{1}{c|}{3, 3} \\ \cline{2-3} 
\end{tabular}
\caption{Symmetric Chicken}
\label{game:symmetric_chicken}
\end{table}
\end{toappendix}

For example, consider Symmetric Chicken (\Cref{game:symmetric_chicken} in the
appendix), which is identical to Asymmetric Chicken
(\Cref{game:asymmetric_chicken_and_p_token_spi}) except that the payoffs for
(Dare, Swerve) are $(5,1)$ rather than $(6,1)$.
Every outcome aside from (Dare, Dare) has social welfare $6$, so there is no
payments token game SPI.
However, there is an outside bettors token SPI as long as some bettor pays a strictly
positive price for the security, which is worth $6 - (-18) = 24$ if (Dare, Dare)
occurs and $0$ otherwise.
Any bettor who believes (Dare, Dare) has nonzero probability should be willing to
pay a strictly positive amount.
For instance, if the bettor expects the players to play the mixed Nash
equilibrium, where each plays Dare with probability $1/6$, the security is worth
$24 \cdot (1/6)^2 = 2/3$ to them.
If they pay this full price, the players' expected social welfare (assuming they do in fact play this mixed Nash) rises from
$16/3$ to exactly $6$, the maximum in Symmetric Chicken---an instance of the
general result we give in \Cref{thm:outside_bettors_SPI_EV}.

Since the security has no downside, even a very risk-averse or uncertain outside
bettor should pay more than $0$ for it, yielding an SPI, unless they are certain it is worthless.

\begin{theoremrep}\label{thm:outside_bettors_SPI_existence2}
    Suppose there is an outside bettor who assigns strictly positive probability to
    exactly the outcomes in $\red A$, and who pays a strictly positive amount for a
    security if and only if it has positive expected value under those beliefs.
    Then an outside bettors token SPI on $G$ fails to exist if and only if
    $\red G$ is constant sum and all outcomes in $A - \red A$ have weakly
    lower social welfare than those in $\red A$.
\end{theoremrep}
\begin{proof}
\textit{Only If:}
Suppose the conditions do not both hold.
If $\red{G}$ is not constant sum, some $a \in \red{A}$ has $\mathrm{SW}(a) < \mathrm{SW}^{max}(G)$. 
If $\red{G}$ is constant sum but some outcome in $A \setminus \red{A}$ has strictly higher social welfare, then every $a \in \red{A}$ satisfies $\mathrm{SW}(a) < \mathrm{SW}^{max}(G)$. 
In either case, an outcome with sub-maximal social welfare exists in the reduced game $\red A$, and hence there is a token outcome $t$ for which the security has positive value when $t$ obtains. 
All outcomes in $\bar{A}$ are possible under our standard assumptions, since all survive iterated elimination of strictly dominated strategies. 
Therefore, the security has strictly positive expected value and sells for a strictly positive price, yielding an SPI.

\textit{If:}
Conversely, suppose $\red{G}$ is constant sum and all outcomes in $A \setminus \red A$ have weakly lower social welfare than those in $\red A$. 
In this case, all token outcomes with positive probability under the standard assumptions achieve the maximum social welfare in $G$, and hence the security pays out $0$ on all such outcomes. 
Therefore, the security has $0$ expected value to a bettor who only assigns positive probability to outcomes possible under the assumptions, and no SPI exists.
\end{proof}

The above characterizes existence, but says nothing about the size of the SPIs.
We can say more if the bettors pay the security's full expected value under some
distribution over token outcomes, as they must if, say, several bettors with
identical beliefs bid for it in a second-price auction.

\begin{theoremrep}\label{thm:outside_bettors_SPI_EV}
    Suppose an outside bettor with probability distribution $D$ over token outcomes purchases the security for its expected value under $D$, and that this value is strictly positive. Then there is an outside bettors token SPI on $G$ in which the players' expected social welfare under $D$ is equal to their maximum social welfare in $G$. 
\end{theoremrep}
\begin{proof}
    Since the security sells for a strictly positive price, the resulting outside bettors token game is an SPI on $G$.
    The players' (total) expected payment to the outside bettor under $D$ is equal to the amount they sell the security for. 
    In other words, the expected net payment between the players and the bettor, again under $D$, is $0$. 
    Therefore, the players' expected social welfare is equal to the (expected) social welfare of the strategy profiles used to realize the token outcomes, and by construction these strategy profiles are social welfare maximizing in the original game $G$.

\end{proof}

Note that the scheme is \textit{not} an SPI for the outside bettor, but an ordinary probabilistic bet. In fact, it is the reverse of a typical insurance contract: the players receive a
fixed amount up front and then pay out in socially bad outcomes. 
So while the players need no probabilistic beliefs, the bettor does. 
We see this as unproblematic: outside parties have no inherent stake in or control over the outcome, and so do not face the incentive issues the players do.

Finally, we note an alternate version of the scheme that works even when no outside party is willing to assign probabilities to outcomes, provided there are multiple sets of players playing disjoint games. 
In this case, the players could jointly agree to \enquote{swap} their securities without any other compensation, i.e. players from game $1$ commit to transfer the surplus from their game to the players in game $2$, and vice versa. 
This results in an SPI for all players, without requiring any probabilistic beliefs about the outcomes of the games. 
Each group's gain is now random rather than fixed, but the SPI is still strict as long as the other group's players might not reach a welfare-maximizing outcome.

\section{Computational Experiments}\label{Sec:experimental_results}

\begin{figure}[t]
    \centering
    \includegraphics[width=\linewidth]
    {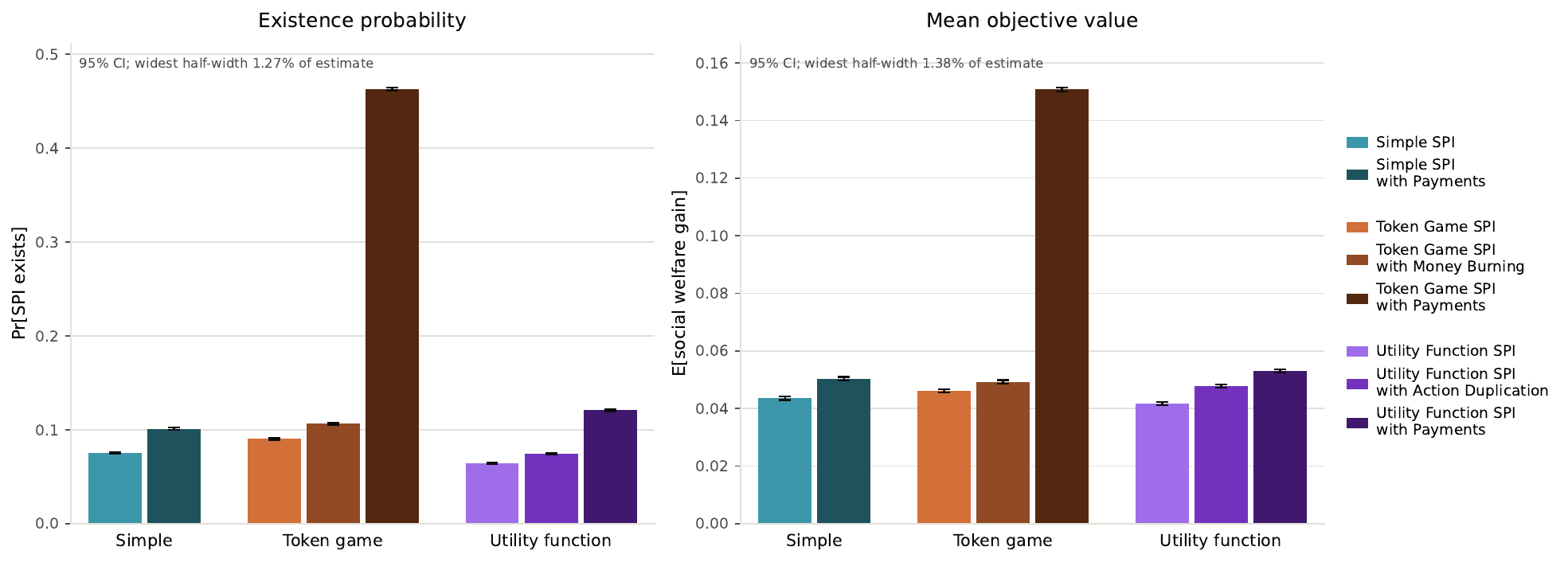}
    \caption{Empirical existence probability and mean objective value of SPIs under %
    a uniform distribution over payoffs.}
    \label{fig:existence_prob_unif}
\end{figure}

We think the interventions we propose in this paper are particularly suitable for achieving SPIs on many strategic interactions.
For outside bettors token games, strong theoretical results support this claim. 
To test our other types of SPIs, we evaluate against existing SPI frameworks on randomly generated games, restricting to $2$ players for simplicity.

\textbf{Experimental setup:}
We generate $50000$ random games of size $m$ by $m$, for each $m \in \{3, \dots, 9\}$. Each entry of the payoff matrix is sampled i.i.d. from the uniform distribution on the interval $[0, 1]$. 
We use iterated elimination of strictly dominated strategies as the reduction function. 
We compute SPIs with respect to three families of frameworks: simple SPIs, token SPIs, and utility function SPIs. 
We consider simple SPIs implemented by a correlated strategy profile and by a correlated strategy profile with payments. %
For token SPIs, we consider plain, MB, and payments token SPIs. 
Finally, we consider utility function SPIs as in \citet{SPI-original}, as well as two augmentations of their framework: action duplication and payments, all of which we explain in detail in \Cref{appendix_subsec:uf_spi_definitions}.  
For every game and every type of SPI we record whether an SPI exists and the optimal SPI's objective value: its expected social welfare gain under the uniform distribution over the reduced default game.
We give further details on our computational setup in \Cref{appendix_subsec:experiment_details}.

\begin{toappendix}
    \begin{figure}[t]
    \centering
    \includegraphics[width=\linewidth]{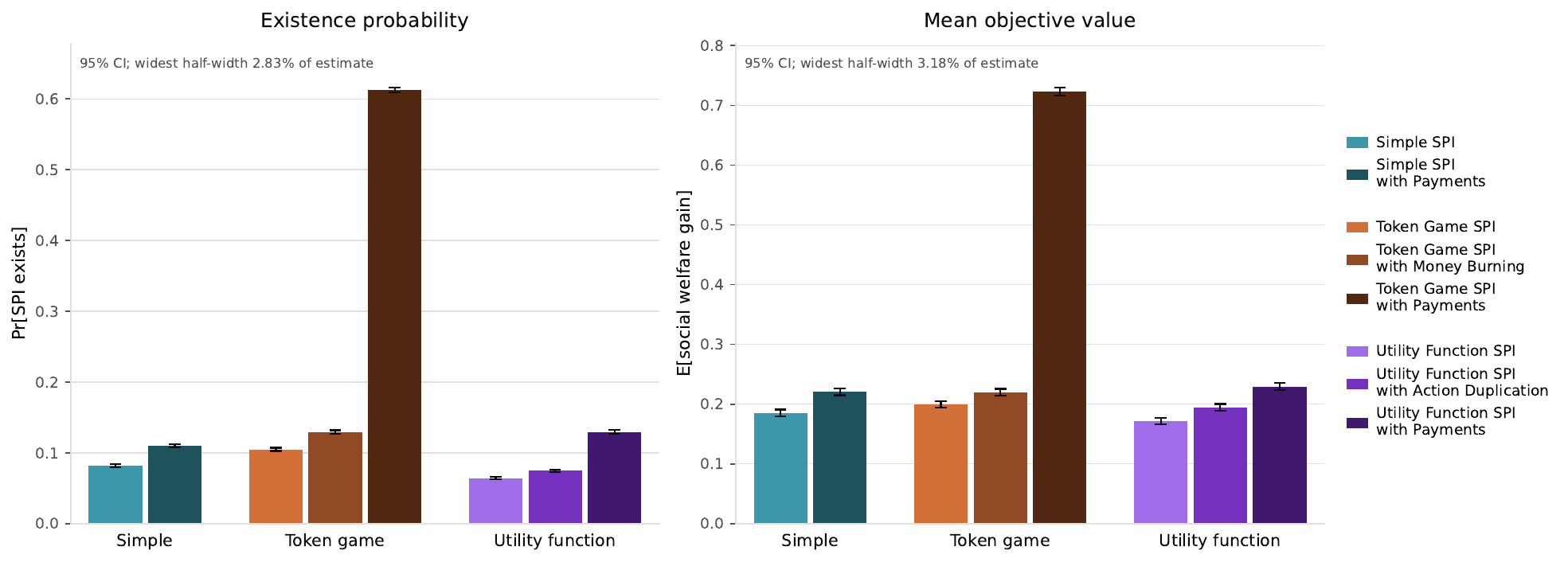}
    \caption{Existence probability and mean objective value with payoffs drawn iid from $N(0,1)$ instead of $U[0,1)$, over 70{,}000 games of sizes 3--9 under pure dominance; otherwise identical to \Cref{fig:existence_prob_unif}. The change of generator leaves the picture qualitatively intact. Objective values are relatively larger, which is intuitive because the distribution has higher variance. 
    }
    \label{existence_and_objective_normal_pure}
\end{figure}

    \begin{figure}[t]
    \centering
    \includegraphics[width=\linewidth]{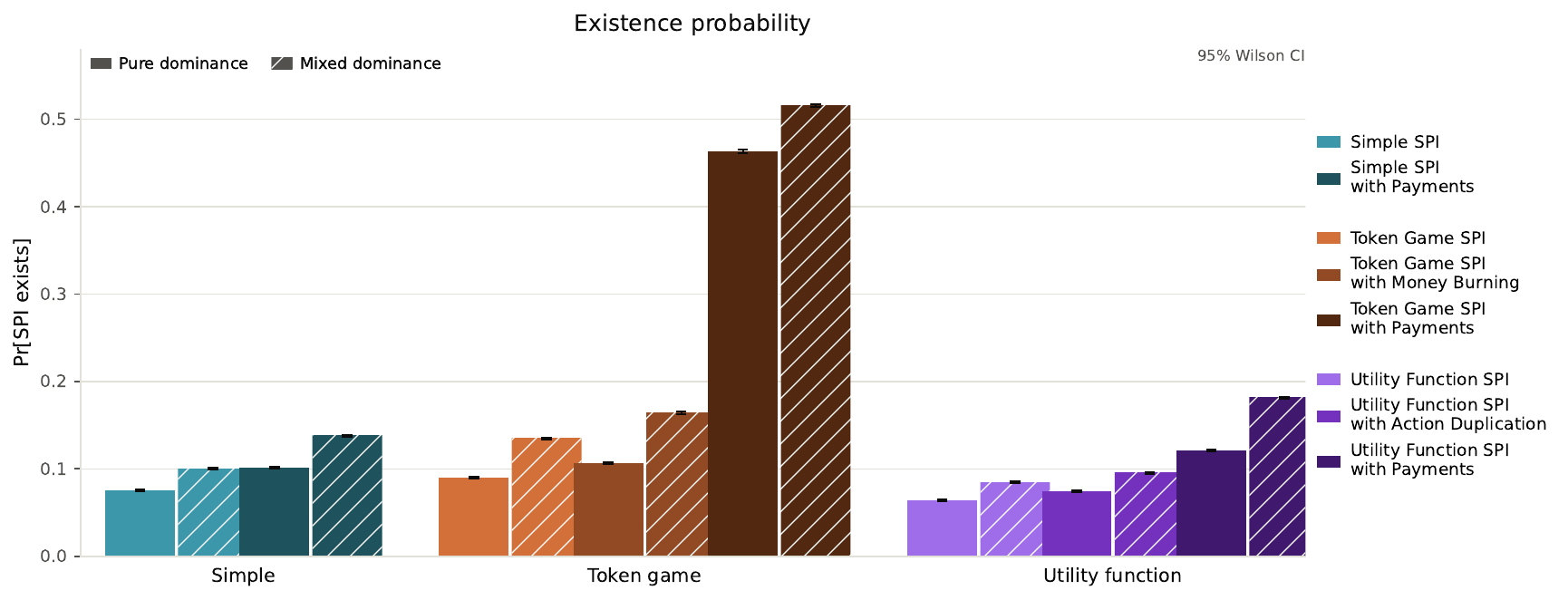}
    \caption{Effect of replacing iterated elimination by \emph{pure} dominance with elimination by \emph{mixed} dominance on the probability that an SPI exists in the standard dataset.
    The comparison is on the same set of games. 
    See Table~\ref{pure_vs_mixed_uniform} for the data and confidence intervals.
    }
    \label{existence_pure_vs_mixed_uniform}
\end{figure}

    \begin{figure}[t]
    \centering
    \includegraphics[width=\linewidth]{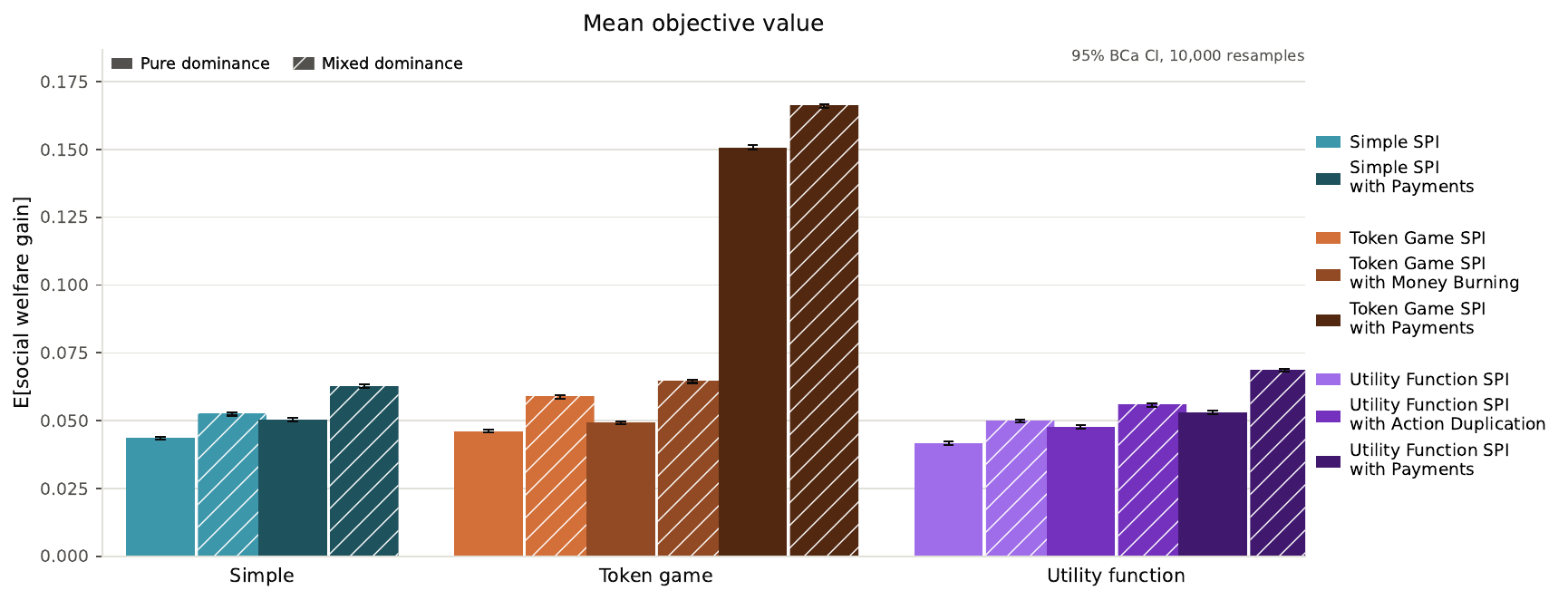}
    \caption{
    Effect of replacing iterated elimination by \emph{pure} dominance with elimination by \emph{mixed} dominance on mean objective value in the standard dataset.
    The comparison is on the same set of games. 
    See Table~\ref{pure_vs_mixed_uniform} for the data and confidence intervals.
    }
    \label{objective_pure_vs_mixed_uniform}
\end{figure}

    \begin{table}[t]
  \centering
  \footnotesize
  \setlength{\tabcolsep}{4pt}
  \resizebox{\textwidth}{!}{%
\begin{tabular}{lcccccc}
    \toprule
    & \multicolumn{3}{c}{Existence probability} & \multicolumn{3}{c}{Mean objective value} \\
    \cmidrule(lr){2-4}\cmidrule(lr){5-7}
    & pure & mixed & increase & pure & mixed & increase \\
    \midrule
    \texttt{simple} & \shortstack{0.0751\\{\scriptsize [0.0742, 0.0760]}} & \shortstack{0.1000\\{\scriptsize [0.0991, 0.1010]}} & \shortstack{+33.25\%\\{\scriptsize [+32.38, +34.21]}} & \shortstack{0.0435\\{\scriptsize [0.0430, 0.0441]}} & \shortstack{0.0525\\{\scriptsize [0.0519, 0.0531]}} & \shortstack{+20.55\%\\{\scriptsize [+19.69, +21.48]}} \\
    \texttt{simple\_payments} & \shortstack{0.1012\\{\scriptsize [0.1002, 0.1022]}} & \shortstack{0.1379\\{\scriptsize [0.1368, 0.1391]}} & \shortstack{+36.35\%\\{\scriptsize [+35.58, +37.18]}} & \shortstack{0.0504\\{\scriptsize [0.0498, 0.0510]}} & \shortstack{0.0627\\{\scriptsize [0.0621, 0.0633]}} & \shortstack{+24.40\%\\{\scriptsize [+23.57, +25.28]}} \\
    \texttt{token} & \shortstack{0.0903\\{\scriptsize [0.0894, 0.0913]}} & \shortstack{0.1348\\{\scriptsize [0.1336, 0.1359]}} & \shortstack{+49.15\%\\{\scriptsize [+48.14, +50.19]}} & \shortstack{0.0461\\{\scriptsize [0.0456, 0.0467]}} & \shortstack{0.0588\\{\scriptsize [0.0582, 0.0594]}} & \shortstack{+27.48\%\\{\scriptsize [+26.59, +28.44]}} \\
    \texttt{token\_mb} & \shortstack{0.1064\\{\scriptsize [0.1054, 0.1074]}} & \shortstack{0.1639\\{\scriptsize [0.1627, 0.1652]}} & \shortstack{+54.07\%\\{\scriptsize [+53.07, +55.08]}} & \shortstack{0.0492\\{\scriptsize [0.0486, 0.0498]}} & \shortstack{0.0645\\{\scriptsize [0.0639, 0.0652]}} & \shortstack{+31.10\%\\{\scriptsize [+30.22, +32.03]}} \\
    \texttt{token\_payments} & \shortstack{0.4632\\{\scriptsize [0.4616, 0.4649]}} & \shortstack{0.5158\\{\scriptsize [0.5141, 0.5175]}} & \shortstack{+11.35\%\\{\scriptsize [+11.13, +11.58]}} & \shortstack{0.1508\\{\scriptsize [0.1501, 0.1515]}} & \shortstack{0.1660\\{\scriptsize [0.1653, 0.1668]}} & \shortstack{+10.11\%\\{\scriptsize [+9.83, +10.40]}} \\
    \texttt{uf\_spi} & \shortstack{0.0641\\{\scriptsize [0.0633, 0.0649]}} & \shortstack{0.0846\\{\scriptsize [0.0837, 0.0855]}} & \shortstack{+32.00\%\\{\scriptsize [+31.06, +33.02]}} & \shortstack{0.0417\\{\scriptsize [0.0411, 0.0423]}} & \shortstack{0.0499\\{\scriptsize [0.0493, 0.0505]}} & \shortstack{+19.57\%\\{\scriptsize [+18.68, +20.53]}} \\
    \texttt{uf\_spi\_ad} & \shortstack{0.0744\\{\scriptsize [0.0736, 0.0753]}} & \shortstack{0.0952\\{\scriptsize [0.0943, 0.0962]}} & \shortstack{+27.95\%\\{\scriptsize [+27.10, +28.87]}} & \shortstack{0.0478\\{\scriptsize [0.0472, 0.0484]}} & \shortstack{0.0558\\{\scriptsize [0.0552, 0.0564]}} & \shortstack{+16.80\%\\{\scriptsize [+15.98, +17.65]}} \\
    \texttt{uf\_spi\_payments} & \shortstack{0.1210\\{\scriptsize [0.1199, 0.1221]}} & \shortstack{0.1812\\{\scriptsize [0.1799, 0.1825]}} & \shortstack{+49.80\%\\{\scriptsize [+48.91, +50.72]}} & \shortstack{0.0530\\{\scriptsize [0.0524, 0.0536]}} & \shortstack{0.0686\\{\scriptsize [0.0679, 0.0692]}} & \shortstack{+29.45\%\\{\scriptsize [+28.59, +30.34]}} \\
    \bottomrule
\end{tabular}
  }
  \caption{
  Existence probability and objective values for each SPI framework on the standard dataset under both pure and mixed dominance.
  Existence columns have 95\% Wilson score intervals. 
  Objective columns have BCa bootstrap confidence intervals, 10,000 resamples.
  For the increase columns, each percent increase and its 95\% interval come from a paired BCa bootstrap that resamples games and recomputes both arms per resample. 
  Data for \Cref{fig:existence_prob_unif,existence_pure_vs_mixed_uniform,objective_pure_vs_mixed_uniform}.
  }
  \label{pure_vs_mixed_uniform}
\end{table}

\end{toappendix}

\textbf{Basic Results:}
\Cref{fig:existence_prob_unif} shows the empirical existence probability and mean social welfare gains under each SPI framework. 
We also find qualitatively similar results when sampling game payoffs from a standard normal distribution (see \Cref{existence_and_objective_normal_pure} in the appendix). 
The interventions we introduce in this paper, token games with money burning and token games with payments, are substantially more effective than each of plain token SPIs, utility function SPIs, and the baseline of simple SPIs. 
The combination of token games and payments is especially potent: payments improve token SPIs far more than they
improve simple or utility function SPIs.

\textbf{Mixed Dominance:}
While all other results use pure dominance, we here consider using a stronger mixed dominance assumption: iterated elimination of actions strictly dominated by mixed rather than pure strategies.  
Across frameworks, switching to mixed dominance provides a relative increase in SPI existence probability of between $11.3\%$ 
and $53.4\%$%
~(\Cref{existence_pure_vs_mixed_uniform}). See \Cref{appendix_subsec:mixed_dominance} for more detail. 
\begin{toappendix}
\subsection{Further Details on Mixed Dominance}\label{appendix_subsec:mixed_dominance}
While it's intuitive that mixed dominance enables more SPIs, it's not immediately implied by the fact that mixed dominance is stronger. 
This is because SPIs must be \textit{strictly} Pareto improving for at least one outcome in the reduced game, so eliminating a weak outcome can make achieving SPIs more difficult. 
For instance, suppose there's a game which reduces to a single outcome under mixed dominance, and this outcome is each player's unique maximum utility in the full game.
Such a game admits no SPI under mixed dominance, but if pure dominance is unable to reduce it to the single outcome, then it admits a (simple) SPI under pure dominance.
\end{toappendix}

\textbf{Evaluating the Induced SPI Selection Problem:}
Thus far, we've been concerned only with the existence and magnitude of SPIs. 
But multiple SPIs often exist on the same game, and for SPIs to succeed in benefiting the players, the players must be able to agree on a particular SPI to play. 
\citet[][Section 6]{SPI-original} call this the SPI selection problem. 
Hence, it's important to evaluate SPI frameworks on the difficulty of the SPI selection problem they induce. 

For a given game and SPI framework, we consider the bargaining problem whose feasible region is the possible SPIs, plus the disagreement point of playing the base game directly. 
We analyze each framework's frequency of producing genuine bargaining instances, as opposed to ones where the players' favorite SPIs coincide or one player's value is zero across all SPIs. 
Among genuine bargaining instances, we further analyze the fractions of each player's maximum gain they achieve at the other player's favorite SPI and at the \citet{nash1950bargaining} bargaining solution. 

Overall, our results suggest that utility function SPIs tend to induce easier bargaining problems than the other SPI families, and that within each SPI framework, allowing payments tends to result in bargaining problems that are more difficult, which makes sense given that payments enable zero-sum redistribution of gains.
See \Cref{subsec:SPI_selection} for further detail.

\begin{toappendix}
\subsection{Evaluating the Induced SPI Selection Problem}
\label{subsec:SPI_selection}
Thus far, we've been concerned only with the existence and magnitude of SPIs. 
But multiple SPIs often exist on the same game, and for SPIs to succeed in benefiting the players, the players must be able to agree on a particular SPI to play. 
\citet[][Section 6]{SPI-original} call this the SPI selection problem. 
An additional important dimension for evaluating SPI frameworks is the difficulty of the SPI selection problem they induce. 

We will think of the set of available SPIs as inducing a bargaining problem over SPIs with a disagreement point of playing the base game directly.  
We'll assume the value of each SPI for each player is the player's expected utility gain under the uniform distribution over reduced game outcomes.  
As before, this is a rough proxy for the value of each SPI, which depends on the unknown true distribution over outcomes. 
Hence, the disagreement point has value $(0,0)$. 

Overall, our results suggest that utility function SPIs tend to induce easier bargaining problems than the other SPI families, and that within each SPI framework, allowing payments tends to result in bargaining problems that are more difficult, which makes sense given that payments enable zero-sum redistribution of gains.

First, we analyze whether the various frameworks produce genuine bargaining problems at all. 
We categorize the bargaining instance produced by each game and SPI framework into one of 4 categories: 
no SPI exists, strict SPIs exist only for one player, the players' favorite SPIs coincide, and genuine bargaining instances (i.e., none of the prior three). 

Only the token SPI family ever falls into the one-sided regime. 
We can also see that this is what causes token SPIs with payments to be so dominant in terms of existence probability (one-sided SPIs in $33\%$ of all games, which constitute $71\%$ of its SPIs). 
Games where only one-sided simple or utility function SPIs can exist would require exact payoff equalities, which occur with probability zero over our uniform payoff distribution.
Given that both players can achieve strict SPIs, the probabilities of the favorites coinciding fall into three tiers: the two types of utility function SPIs without payments are highest by far, token and simple SPIs without payments are moderate, and the frameworks with payments are lowest.
See \Cref{shape_composition_strict_for_both} and \Cref{regime_shares} for more detail. 

Then, we focus on the subset of genuine bargaining instances and analyze their difficulty. 
For this, we measure two statistics, both of which are fractions of a player's maximum gain over SPIs. 
First, the concession ratio, which is the maximum gain ratio Player $1$ can achieve given Player $2$ achieves their maximum possible gain over SPIs. 
In other words, this is Player $1$'s gain ratio at Player $2$'s favorite (Pareto optimal) SPI. 
Second, we report the (Nash) compromise ratio, which is the gain ratio a player achieves at the Nash bargaining solution. 
We report both averaged over both players and games.
The findings are similar to the probabilities that the players' favorite SPIs coincide: 
Within each SPI family, enabling payments causes both the concession ratio and compromise ratio to drop substantially, and the non-payment utility function SPIs have substantially higher ratios than the non-payment simple and token SPIs. 
(We found very similar results when defining the compromise ratios with respect to other standard bargaining solutions: Kalai-Smorodinsky \cite{kalai1975other_KS-bargaining}, egalitarian \cite{kalai1977proportional_egal_bargaining}, and utilitarian, i.e., social welfare maximizing with ties broken to maximize the minimum gain.)
See \Cref{bargaining_difficulty} for more detail.  

Note that there are further difficulties with bargaining over SPIs that this analysis neglects. 
For example, here we assume that the value of each SPI for each player is common knowledge, but in practice the value of an SPI to a player depends on that player's subjective beliefs about the game's distribution over outcomes.
In addition to being private information, eliciting such beliefs could be distorted by the types of signaling effects discussed in the intro. 
Our aim in the present section is merely to get a rough sense of the structure of bargaining problems induced by each framework.
\end{toappendix}

\begin{toappendix}

\begin{figure}[t]
    \centering
    \includegraphics[width=\linewidth]{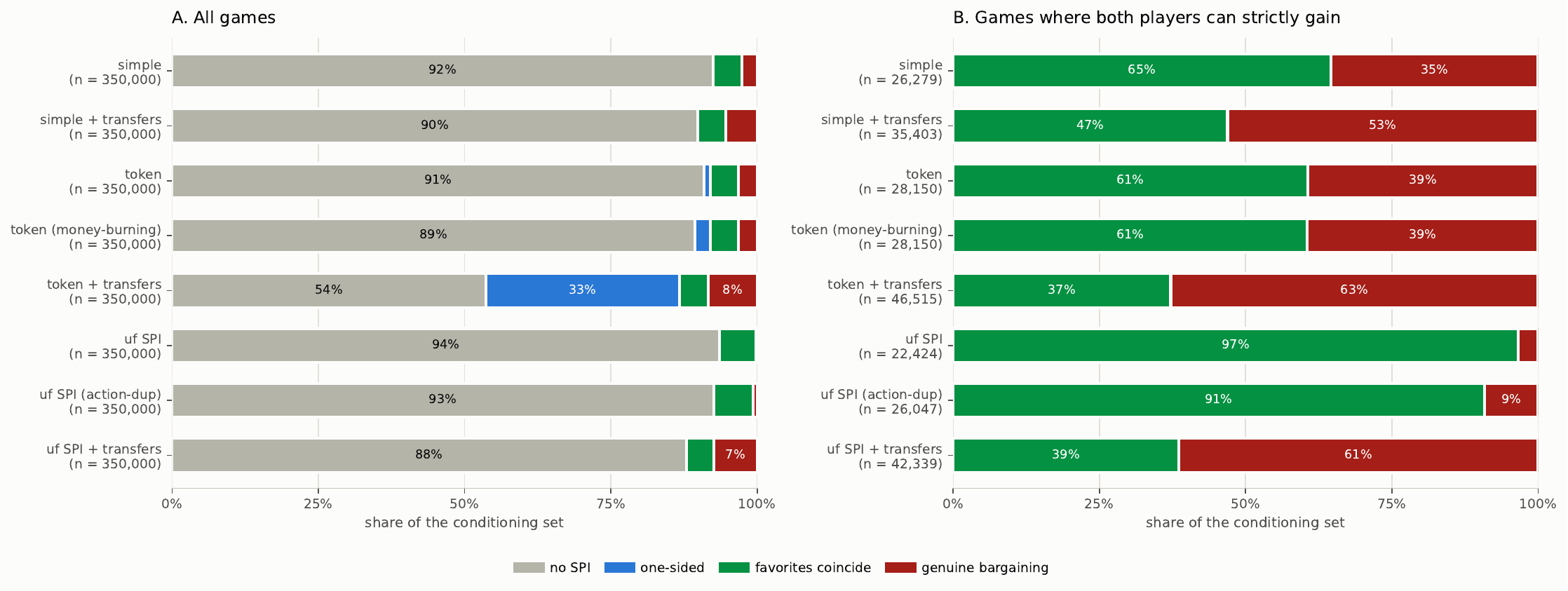}
    \caption{Distribution over bargaining regimes for each SPI framework in the standard dataset. One-sided means a strict SPI exists for only one player. B shows the two-way split conditional on both players having strict SPIs. The data, including confidence intervals, are shown in Table~\ref{regime_shares}.
    }
    \label{shape_composition_strict_for_both}
\end{figure}

\begin{table}[t]
\centering
\caption{Regime induced by each SPI framework, over the standard dataset (50,000 random games per size, sizes 3--9, uniform payoffs, pure dominance). The second line of each row is the 95\% Wilson interval as a signed relative deviation from the estimate. %
}
\label{regime_shares}
\begin{tabular}{l rrrr}
\toprule
Framework & No SPI & One-sided & Favorites coincide & Genuine bargaining \\
\midrule
simple & 92.49 & 0 & 4.85 & 2.66 \\
& {\scriptsize $[-0.1,+0.1]$} & {\scriptsize --} & {\scriptsize $[-1.5,+1.5]$} & {\scriptsize $[-2.0,+2.0]$} \\[2pt]
simple + transfers & 89.88 & 0 & 4.74 & 5.37 \\
& {\scriptsize $[-0.1,+0.1]$} & {\scriptsize --} & {\scriptsize $[-1.5,+1.5]$} & {\scriptsize $[-1.4,+1.4]$} \\[2pt]
token & 90.97 & 0.99 & 4.88 & 3.17 \\
& {\scriptsize $[-0.1,+0.1]$} & {\scriptsize $[-3.3,+3.4]$} & {\scriptsize $[-1.5,+1.5]$} & {\scriptsize $[-1.8,+1.8]$} \\[2pt]
token (money-burning) & 89.36 & 2.60 & 4.87 & 3.17 \\
& {\scriptsize $[-0.1,+0.1]$} & {\scriptsize $[-2.0,+2.0]$} & {\scriptsize $[-1.5,+1.5]$} & {\scriptsize $[-1.8,+1.8]$} \\[2pt]
token + transfers & 53.68 & 33.03 & 4.94 & 8.35 \\
& {\scriptsize $[-0.3,+0.3]$} & {\scriptsize $[-0.5,+0.5]$} & {\scriptsize $[-1.4,+1.5]$} & {\scriptsize $[-1.1,+1.1]$} \\[2pt]
uf SPI & 93.59 & 0 & 6.19 & 0.22 \\
& {\scriptsize $[-0.1,+0.1]$} & {\scriptsize --} & {\scriptsize $[-1.3,+1.3]$} & {\scriptsize $[-6.9,+7.4]$} \\[2pt]
uf SPI (action-dup) & 92.56 & 0 & 6.76 & 0.68 \\
& {\scriptsize $[-0.1,+0.1]$} & {\scriptsize --} & {\scriptsize $[-1.2,+1.2]$} & {\scriptsize $[-3.9,+4.1]$} \\[2pt]
uf SPI + transfers & 87.90 & 0 & 4.67 & 7.43 \\
& {\scriptsize $[-0.1,+0.1]$} & {\scriptsize --} & {\scriptsize $[-1.5,+1.5]$} & {\scriptsize $[-1.2,+1.2]$} \\[2pt]
\bottomrule
\end{tabular}
\end{table}

\begin{table}[t]
\centering
\providecommand{\SpiExpect}{\mathrm{E}}
\caption{Bargaining difficulty conditional on genuine bargaining, averaged over the two players. $s_i(1)$ is the concession ratio: the maximum gain ratio (relative to their maximum gain over SPIs) one player can achieve given the other player achieves their maximum gain. 
$s_i(\mathrm{NBS})$ is the (Nash) compromise ratio: the gain ratio a player achieves at the Nash bargaining solution.  
Brackets are 95\% BCa intervals as signed relative deviations from the estimate, bootstrapped over games. 
$\bar \Delta$ is the mean over games of the spread (maximum absolute difference) between the players' average $s_i(b)$ computed at the $4$ bargaining solutions. 
$n$ is the number of genuine-bargaining games, out of 350{,}000.}
\label{bargaining_difficulty}
\begin{tabular}{l r rr r}
\toprule
Framework & $n$ & $\mathrm{E}[s_i(1)]$ & $\mathrm{E}[s_i(\mathrm{NBS})]$ & $\bar{\Delta}$ \\
\midrule
simple & 9{,}305 & 0.139 & 0.666 & 0.058 \\
& & {\scriptsize $[-3.4,+3.4]$} & {\scriptsize $[-0.5,+0.6]$} & \\[2pt]
simple + transfers & 18{,}800 & 0.099 & 0.552 & 0.001 \\
& & {\scriptsize $[-3.7,+3.9]$} & {\scriptsize $[-0.3,+0.4]$} & \\[2pt]
token & 11{,}084 & 0.149 & 0.683 & 0.063 \\
& & {\scriptsize $[-2.8,+2.8]$} & {\scriptsize $[-0.5,+0.5]$} & \\[2pt]
token (money-burning) & 11{,}108 & 0.141 & 0.680 & 0.061 \\
& & {\scriptsize $[-2.8,+2.8]$} & {\scriptsize $[-0.5,+0.5]$} & \\[2pt]
token + transfers & 29{,}209 & 0.086 & 0.617 & 0.023 \\
& & {\scriptsize $[-2.6,+2.6]$} & {\scriptsize $[-0.3,+0.3]$} & \\[2pt]
uf SPI & 759 & 0.462 & 0.818 & 0.046 \\
& & {\scriptsize $[-3.0,+3.1]$} & {\scriptsize $[-1.1,+1.0]$} & \\[2pt]
uf SPI (action-dup) & 2{,}386 & 0.754 & 0.922 & 0.025 \\
& & {\scriptsize $[-1.3,+1.3]$} & {\scriptsize $[-0.5,+0.4]$} & \\[2pt]
uf SPI + transfers & 25{,}995 & 0.067 & 0.539 & 0.001 \\
& & {\scriptsize $[-3.8,+4.0]$} & {\scriptsize $[-0.3,+0.3]$} & \\[2pt]
\bottomrule
\end{tabular}
\end{table}

\end{toappendix}

\begin{toappendix}
    \subsection{Details of Computational Experiments}
    \label{appendix_subsec:experiment_details}

    For the different types of token SPIs we implement the linear programs of \Cref{Sec:mb-token-SPIs,Sec:payment-token-SPIs} and \citet[Section 4]{Sauerberg2025_promises_EPV-SPIs}. 
    For iterated elimination of strictly dominated strategies in mixed strategies we also use linear programs \cite{Conitzer2005}. 
    For linear programming, we use the CVXPY modeling language \citep{diamond2016cvxpy} and the MOSEK LP solver \cite{mosek}.
    Computing utility function SPIs is NP-complete \cite{SPI-original}. However, using a classic depth-first search algorithm %
    we are able to find (and optimize over) utility function SPIs in acceptable time for the given game sizes.

    The minimum game size of $3 \times 3$ was chosen because for $2 \times 2$ games (with continuous payoffs) if the game is reducible it will reduce down to a single outcome, which renders the simple and token SPI frameworks equivalent. 
    The maximum game size of $9 \times 9$ was chosen essentially arbitrarily, but we chose not to do much larger sizes both because of time constraints and because at larger game sizes, fewer games are reducible and hence the results seem less interesting. 

    Finally, we have two important hyperparameters used to deal with tolerances to LP solver imprecision and floating point arithmetic errors. 
    We used $10^{-9}$ as our threshold for the objective value required for an SPI to be considered strictly Pareto improving and hence valid. 
    On test runs, this was well above the maximum objective value the LPs returned on games in which no SPI exists, so the true value should be $0$. 
    The smallest SPIs found during test runs had objective values much larger than this threshold and the smallest few were manually verified to be genuine.
    The solver errors can be much larger when reading the values of variables at the optimal solution; MOSEK allows constraints to be violated by $10^{-8}$. 
    Because of this, we used a tolerance of $10^{-5}$ when comparing variable values for equality, which was found to be satisfactory on test runs. 
    
    Our main data collection run was done locally on a laptop and took about 17 hours. 

    Throughout, the standard dataset refers to the dataset of 50,000 games of each size from $3 \times 3$ to $9 \times 9$ with payoffs drawn uniformly from $[0,1]$.
    
\end{toappendix}

\begin{toappendix}
    \subsection{Definitions for Other SPI Frameworks}
    \label{appendix_subsec:uf_spi_definitions}

    \subsubsection{Utility Function SPIs}

    Utility function SPIs \cite{SPI-original} are SPIs achieved by what we'll call utility function commitment, which \citet{SPI-original} introduce in a delegation framework. 
    The idea is that the players will delegate to some representatives to play on their behalf, by default telling them to play the original game with the players' true utility functions. 
    But the players can also (jointly) instruct their agents to play a subgame $G'$ of $G$ according to a utility function which is different from the players' true utility function on this subgame. 
    In particular, if the delegates are instructed to play a subgame of the same shape (number of actions per player) as $\red G$ and according to the players' utilities on $\red G$, then the subgame's outcome will be (isomorphic to) whatever the delegates would have played in $G$ by default. 
    The result is that a utility function SPI on $G$ exists if and only if there is a subgame $G'$ of $G$ and player-wise bijection $\{\iso_i\ :  \red A_i \rightarrow A'_i\}$ from the reduced game $\red G$ to $G'$ for which $\iso(a)$ Pareto dominates $a$ for all $a \in \red A$, and strictly so for at least one $a \in \red A$. 
    Roughly, an SPI exists if there's a subgame of $G$ of the same shape as $\red G$ but outcome-wise Pareto better. 
    See \citet{SPI-original} for a more formal treatment.

    In addition, we consider two extensions of the utility function SPI framework. 

    First, we consider utility function SPIs with action duplication.
    This allows the players to duplicate actions in $G$ before constructing the subgame $G'$: $G'$ can contain multiple \enquote{copies} of the same action in $G$ (which is useful because the representatives' utility functions can treat the different copies as distinct actions with different payoffs).
    The characterization of SPIs under this form of commitment is identical to the above except that rather than a subgame of $G$, $G'$ is a subgame of $G^{dup}$: a version of $G$ with $|A_i|$ copies of Player $i$'s actions, for each $i$. 

    Second, we consider utility function SPIs with payments. 
    The payments are analogous to those in \Cref{Sec:payment-token-SPIs}: utility transfers between the players that can condition on  the outcome of the representatives' game. 
    The result is again a similar characterization, except that corresponding outcomes must be social welfare improving rather than Pareto improving: 
    There is an SPI if and only if there is a subgame $G'$ of $G$ and player-wise bijection $\{\iso_i\ :  \red A_i \rightarrow A'_i\}$ from the reduced game $\red G$ to $G'$ for which $\SW(\iso(a)) \geq \SW(a)$ for all $a \in \red A$, and strictly so for at least one $a \in \red A$.

    \subsubsection{Simple SPIs}

    The \enquote{simple SPIs} baseline consists of the SPIs that can be achieved by commitment to a single strategy profile, and simple SPIs with payments are those achieved by commitment to a single strategy profile and payment profile.
    This terminology is slightly inconsistent, because we have no name for these forms of commitment separate from the SPIs they achieve (as we do for token game commitment), and because, e.g., token game SPIs can be either simple or isomorphism SPIs.
    But the point is that these simple commitment frameworks can \textit{only} achieve simple SPIs, in contrast with the more involved frameworks, token game and utility function commitment, which can achieve isomorphism SPIs because the players' payoffs depend on the result of a token game or on the delegates' choices.

\end{toappendix}

\bigskip

\bibliography{references}

\begin{thebibliography}{26}
\providecommand{\natexlab}[1]{#1}

\bibitem[{Anderson, Shoham, and Altman(2010)}]{internal_implementation}
Anderson, A.; Shoham, Y.; and Altman, A. 2010.
\newblock Internal Implementation.
\newblock In \emph{Proceedings of the 9th International Conference on
  Autonomous Agents and Multiagent Systems: Volume 1}, AAMAS '10, 191–198.
  Richland, SC: International Foundation for Autonomous Agents and Multiagent
  Systems.
\newblock ISBN 9780982657119.

\bibitem[{Apt(2004)}]{Apt2004}
Apt, K.~R. 2004.
\newblock Uniform Proofs of Order Independence for Various Strategy Elimination
  Procedures.
\newblock \emph{The B.E.~Journal of Theoretical Economics}, 4(1): 1--48.

\bibitem[{Cavallo(2006)}]{Cavallo2006_redistribution_VCG_payments}
Cavallo, R. 2006.
\newblock Optimal decision-making with minimal waste: strategyproof
  redistribution of {VCG} payments.
\newblock In \emph{Proceedings of the Fifth International Joint Conference on
  Autonomous Agents and Multiagent Systems}, AAMAS '06, 882–889. New York,
  NY, USA: Association for Computing Machinery.
\newblock ISBN 1595933034.

\bibitem[{Conitzer and Sandholm(2005)}]{Conitzer2005}
Conitzer, V.; and Sandholm, T. 2005.
\newblock Complexity of (Iterated) Dominance.
\newblock In \emph{Proceedings of the 6th ACM conference on Electronic
  commerce}, 88--97. Vancouver, Canada: Association for Computing Machinery.

\bibitem[{Deng and Conitzer(2017)}]{deng2017disarmament}
Deng, Y.; and Conitzer, V. 2017.
\newblock Disarmament Games.
\newblock \emph{Proceedings of the AAAI Conference on Artificial Intelligence},
  31(1).

\bibitem[{Diamond and Boyd(2016)}]{diamond2016cvxpy}
Diamond, S.; and Boyd, S. 2016.
\newblock {CVXPY}: A Python-embedded modeling language for convex optimization.
\newblock \emph{Journal of Machine Learning Research}, 17(83): 1--5.

\bibitem[{DiGiovanni, Clifton, and Mac{\'e}(2025)}]{digiovanni2024safe}
DiGiovanni, A.; Clifton, J.; and Mac{\'e}, N. 2025.
\newblock Safe {P}areto Improvements for Expected Utility Maximizers in Program
  Games.
\newblock In \emph{Proceedings of the 24th International Conference on
  Autonomous Agents and Multiagent Systems}.

\bibitem[{Faltings(2004)}]{faltings2004_budget}
Faltings, B. 2004.
\newblock A budget-balanced, incentive-compatible scheme for social choice.
\newblock In \emph{International Workshop on Agent-Mediated Electronic
  Commerce}, 30--43. Springer.

\bibitem[{Gilboa, Kalai, and Zemel(1990)}]{Gilboa1990}
Gilboa, I.; Kalai, E.; and Zemel, E. 1990.
\newblock On the order of eliminating dominated strategies.
\newblock \emph{Operations Research Letters}, 9(2): 85--89.

\bibitem[{Kalai et~al.(2010)Kalai, Kalai, Lehrer, and
  Samet}]{kalai2010commitment}
Kalai, A.~T.; Kalai, E.; Lehrer, E.; and Samet, D. 2010.
\newblock A commitment folk theorem.
\newblock \emph{Games and Economic Behavior}, 69(1): 127--137.

\bibitem[{Kalai(1977)}]{kalai1977proportional_egal_bargaining}
Kalai, E. 1977.
\newblock Proportional Solutions to Bargaining Situations: Interpersonal
  Utility Comparisons.
\newblock \emph{Econometrica}, 45(7): 1623--1630.

\bibitem[{Kalai and Smorodinsky(1975)}]{kalai1975other_KS-bargaining}
Kalai, E.; and Smorodinsky, M. 1975.
\newblock Other Solutions to {Nash's} Bargaining Problem.
\newblock \emph{Econometrica}, 43(3): 513--518.

\bibitem[{Kohlberg and Mertens(1986)}]{Kohlberg1986}
Kohlberg, E.; and Mertens, J.-F. 1986.
\newblock On the Strategic Stability of Equilibria.
\newblock \emph{Econometrica}, 54(5): 1003--1037.

\bibitem[{Monderer and Tennenholtz(2004)}]{k-implementation}
Monderer, D.; and Tennenholtz, M. 2004.
\newblock k-Implementation.
\newblock \emph{Journal of Artificial Intelligence Research}, 21: 37--62.

\bibitem[{{MOSEK ApS}(2025)}]{mosek}
{MOSEK ApS}. 2025.
\newblock \emph{MOSEK Optimizer API for Python 11.2.2}.

\bibitem[{Moulin(1976)}]{moulin1976_money_burning}
Moulin, H. 1976.
\newblock Cooperation in mixed equilibrium.
\newblock \emph{Mathematics of Operations Research}, 1(3): 273--286.

\bibitem[{Moulin(2009)}]{moulin2009almost_BB_VCG_residual_claimant}
Moulin, H. 2009.
\newblock Almost budget-balanced {VCG} mechanisms to assign multiple objects.
\newblock \emph{Journal of Economic Theory}, 144(1): 96--119.

\bibitem[{Nash(1950)}]{nash1950bargaining}
Nash, J.~F., Jr. 1950.
\newblock The Bargaining Problem.
\newblock \emph{Econometrica}, 18(2): 155--162.

\bibitem[{Oesterheld and Conitzer(2022)}]{SPI-original}
Oesterheld, C.; and Conitzer, V. 2022.
\newblock Safe {P}areto improvements for delegated game playing.
\newblock \emph{Autonomous Agents and Multi-Agent Systems}, 36(2): 46.

\bibitem[{Oesterheld and Conitzer(2025)}]{Oesterheld2025}
Oesterheld, C.; and Conitzer, V. 2025.
\newblock Choosing what game to play with no regrets or controversies --
  inferring safe ({P}areto) improvements in binary constraint structures.
\newblock In Bjorndahl, A., ed., \emph{Proceedings of the Twentieth Conference
  on Theoretical Aspects of Rationality and Knowledge (TARK 2025)}, 246--265.
  Düsseldorf, Germany.

\bibitem[{Oesterheld et~al.(2023)Oesterheld, Treutlein, Grosse, Conitzer, and
  Foerster}]{oesterheld2023similarity}
Oesterheld, C.; Treutlein, J.; Grosse, R.~B.; Conitzer, V.; and Foerster, J.
  2023.
\newblock Similarity-based cooperative equilibrium.
\newblock \emph{Advances in Neural Information Processing Systems}, 36:
  24434--24465.

\bibitem[{Pearce(1984)}]{Pearce1984}
Pearce, D.~G. 1984.
\newblock Rationalizable Strategic Behavior and the Problem of Perfection.
\newblock \emph{Econometrica}, 52(4): 1029--1050.

\bibitem[{Sauerberg and Oesterheld(2024)}]{sauerberg2024commitment-to-payments}
Sauerberg, N.; and Oesterheld, C. 2024.
\newblock Computing Optimal Commitments to Strategies and Outcome-Conditional
  Utility Transfers.
\newblock In \emph{Proceedings of the 23rd International Conference on
  Autonomous Agents and Multiagent Systems}, 1654--1663.

\bibitem[{Sauerberg and Oesterheld(2026)}]{Sauerberg2025_promises_EPV-SPIs}
Sauerberg, N.; and Oesterheld, C. 2026.
\newblock Promises Made, Promises Kept: Safe {P}areto Improvements via Ex Post
  Verifiable Commitments.
\newblock In \emph{Proceedings of the AAAI Conference on Artificial
  Intelligence}.

\bibitem[{Souza and R{\^e}go(2012)}]{money-burning-2x2}
Souza, F.; and R{\^e}go, L. 2012.
\newblock Mixed equilibrium: when burning money is rational.

\bibitem[{Tennenholtz(2004)}]{Tennenholtz04:Program}
Tennenholtz, M. 2004.
\newblock Program Equilibrium.
\newblock \emph{Games and Economic Behavior}, 49: 363--373.

\end{thebibliography}

\begin{toappendix}
\section{Addressing the Possible Multiplicity of Isomorphisms}\label{appendix_sec:multiple_isomorphisms}

\begin{lemma}
    Let $G=(A,\mathbf u)$ and $G' = (A', \mathbf u ')$ be two normal-form games. Let $\phi$ and $\hat\phi$ be two isomorphisms from $G$ to $G'$.
    Then for all $a$ in $A$ we have that $\mathbf u ' (\phi(a)) = \mathbf u ' (\hat\phi(a))$.
\end{lemma}

\begin{proofsketch}
    Both isomorphisms must map each player's highest/lowest-utility outcomes in $G$ to that player's highest/lowest-utility outcomes in $G'$. This uniquely determines the coefficients $m_i$ and $b_i$ associated with the isomorphism. Thus all isomorphisms are associated with the same positive affine transformation on payoffs. Since every outcome $a$ has a unique payoff profile, any outcome $a$ must be mapped by all isomorphisms onto outcomes with the same utility in $G'$.
\end{proofsketch}

The following result follows immediately.

\begin{proposition}\label{prop:one-is-pareto-improving-implies-all-pareto-improving}
Let $G$ and $G'$ be two normal-form games. Let $\phi$ and $\hat \phi$ be isomorphisms from $G$ to $G'$. If $\phi$ is Pareto improving, so is $\hat \phi$.
\end{proposition}

This proposition justifies a subtlety of the definition of isomorphism SPIs (\Cref{def:iso-spi}). The definition only requires the existence of a single Pareto-improving isomorphism. Without \Cref{prop:one-is-pareto-improving-implies-all-pareto-improving}, one might wonder what happens if there are multiple isomorphisms, some of which are and some of which aren't Pareto improving. In such a case, \Cref{def:iso-spi} would be questionable. Fortunately, \Cref{prop:one-is-pareto-improving-implies-all-pareto-improving} shows that such cases cannot occur.

\end{toappendix}

\end{document}